\documentclass[12pt,reqno]{article}
\usepackage{amsmath}
\usepackage{amsthm}
\usepackage{graphicx,indentfirst}
\usepackage[psamsfonts]{amssymb}
\numberwithin{equation}{section}

\def\be{\begin{equation}}
\def\ee{\end{equation}}
\def\a{\alpha}

\def \Fig#1#2#3 {
\begin{figure}
\centering
\epsfxsize=#2cm \epsfbox{#1.eps}
\caption{#3}
\label{#1}
\end{figure}
}

\newcount\figno
\def\fig#1#2#3{
\par\begingroup\parindent=0pt\leftskip=1cm\rightskip=1cm\parindent=0pt
\baselineskip=15pt
\global\advance\figno by 1
\epsfxsize=#3
\centerline{\epsfbox{#2}}
\vskip 12pt
{\bf \small Figure \the\figno:} {\small #1}\par
\endgroup\par
}
\def\figlabel#1{\xdef#1{\the\figno
\mbox{ }}}
\def\encadremath#1{\vbox{\hrule\hbox{\vrule\kern8pt\vbox{\kern8pt
\hbox{$\displaystyle #1$}\kern8pt}
\kern8pt\vrule}\hrule}}

\def\b{{\beta}}

\def\bea{\begin{eqnarray*}}
\def\eea{\end{eqnarray*}}

\DeclareMathAlphabet{\mathscr}{OT1}{pzc}{m}{it}

\newcommand{\ud}{\mathrm{d}}
\newcommand{\ub}{\mathrm{b}}
\newcommand{\R}{\mathrm{R}}
\newcommand{\NS}{\mathrm{NS}}

\newcommand{\vm}{{v^{(-)}}}
\newcommand{\vp}{{v^{(+)}}}
\newcommand{\balpha}{{\bar\alpha}}
\newcommand{\bbeta}{{\bar\beta}}

\newcommand{\clebsh}[2]{\left[ {#1 \atop #2} {\alpha_2 \atop x_2} {\alpha_1 \atop x_1} \right]}

\newcommand{\Uqtwo}{\mathcal{U}_q(sl(2))}
\newcommand{\Uqosp}{\mathcal{U}_q(osp(1|2))}
\newcommand{\z}{z}

\newcommand{\chiral}[3]{\!\big[\hspace*{-5pt}\begin{array}{c} \\[-17pt]{\scriptscriptstyle #1}\\[-7pt]{\scriptscriptstyle #2 #3}\end{array}\hspace*{-5pt}\big]}
\title{Self-dual Continuous Series of Representations\\[2mm]
     for $\Uqtwo$ and $\Uqosp$}
\author{Leszek Hadasz$^{a}$, Michal Pawelkiewicz$^b$ and Volker Schomerus$^b$
\\[5mm]
$^a$M. Smoluchowski Institute of Physics, Jagiellonian University,\\
W. Reymonta 4, 30-059 Krak\'ow, Poland
\\[3mm]
$^b$DESY Theory Group, DESY Hamburg  \\
$\,$Notkestrasse 85, D-22603 Hamburg, Germany}

\date{May 2013}

\begin{document}
\begin{titlepage}      \maketitle      \thispagestyle{empty}

\vskip1cm
\begin{abstract}
We determine the Clebsch-Gordan and Racah-Wigner coefficients for continuous
series of representations of the quantum deformed algebras $\Uqtwo$ and
$\Uqosp$. While our results for the former algebra reproduce formulas
by Ponsot and Teschner, the expressions for the orthosymplectic algebra are
new. Up to some normalization factors, the associated Racah-Wigner coefficients
are shown to agree with the fusing matrix in the Neveu-Schwarz sector of $N=1$
supersymmetric Liouville field theory.
\end{abstract}

\vspace*{-16.9cm}\noindent {\tt \phantom{arXiv:yymm.nnnn}}\\ {\tt {DESY 13-083}}
\bigskip\vfill
\noindent
\phantom{wwwx}{\small e-mail: }{\small\tt
\,hadasz@th.if.uj.edu.pl,michal.pawelkiewicz@desy.de,\\
\hspace*{2.3cm} volker.schomerus@desy.de}
\end{titlepage}

\baselineskip=19pt


\setcounter{equation}{0}
\tableofcontents
\newpage

\section{Introduction}

Quantum deformed Lie (super-)algebras are being studied partly because
of their intimate relation with 2-dimensional conformal field theory
and 3-dimensional Chern-Simons theory. Through the investigation of
finite dimensional representations of q-deformed Lie algebras, for
example, one can obtain solutions of the Moore-Seiberg polynomial
equations which describe the fusing and braiding structure of
Wess-Zumino-Witten (WZW) models with compact target group $G$. Because
of the deep relation between WZW models and Chern-Simons theory, the
same data also appear as building blocks for expectation values of
Wilson loops in 3-dimensional topological theories.

The quantum deformed superalgebra $\Uqtwo$ is known to possess a very
interesting self-dual series of infinite dimensional representations,
see e.g. \cite{Schmudgen:1995bx}. As shown by Ponsot and Teschner
\cite{Ponsot:1999uf,Ponsot:2000mt}, this series furnishes a solution of
the Moore-Seiberg relations which is relevant for Liouville conformal field
theory. In fact, the Racah-Wigner coefficients (6j symbols) for these
representations are known to agree with the fusing matrix of Liouville
theory, up to some convention dependent normalization. The same data
also appear in the context of SL(2) Chern-Simons theory or quantum
Teichm\"uller theory \cite{Kashaev:1996kc,Kashaev:1998fc,Teschner:2003em,
Dimofte:2011jd,Andersen:2011bt}. The basic
building block in this case is Faddeev's quantum dilogarithm
\cite{Faddeev:1993rs},
\begin{equation} \label{dilog}
 \Phi_\ub^{+}(z) = \Phi_\ub(z) = \exp\left( \int_C \frac{e^{-2izw}}
{\sinh(w\ub)\sinh(w/\ub)}
\frac{dw}{4w}\right) \ .
\end{equation}
This special function plays an important role in mathematical physics.
It has many beautiful properties, in particular it satisfies a five
term (pentagon) relation \cite{Faddeev:1993rs}. It may be considered
as a quantization of Roger's five-term identity for the ordinary
dilogarithm and can be formulated as an integral identity. The latter
is also known as Ramanujan's identity.

Our goal here is to extend the results described in the
previous paragraph to the Lie superalgebra $\Uqosp$. This algebra
was first introduced in \cite{Kulish:1988gr}. Finite dimensional
representations and their Racah-Wigner coefficients were studied
in \cite{Saleur:1989gj} and other subsequent papers. The series
of representations we are about to analyse below allows us
to obtain the fusing matrix of N=1 Liouville field theory. In
building the basic representation theoretic data, and in
particular the Racah-Wigner coefficients, Faddeev's quantum
dilogarithm gets accompanied by a second function
\begin{equation} \label{dilogminus}
 \Phi_\ub^{-}(z) = \exp\left( \int_C \frac{e^{-2izw}}{\cosh(w\ub)
\cosh(w/\ub)} \frac{dw}{4w}\right) \ .
\end{equation}
From $\Phi_\ub^\pm$ we shall pass to the functions $\Phi^\nu_\ub,
\nu =0,1,$ which are obtained through
\begin{equation} \label{superdilog}
 \Phi^{\nu}_\ub(z) = \log \Phi^-_\ub(z) -
(-1)^\nu \log \Phi^+_\ub(z)\   .
\end{equation}
The precise relation of $\Phi_\ub^{\nu}$ to the functions
$S_\nu$ that appear below is spelled out in Appendix A.2.
Our functions $\Phi^\nu_\ub$ share many features  with
Faddeev's quantum dilogarithm. In particular, they satisfy the
same set of integral identities with an additional sum over
the $\nu$ index wherever the standard identities involve an
integration over $x$. These identities include a new variant
of the pentagon relation for Faddeev's quantum dilogarithm.
Clearly, the pair $\Phi^\nu_\ub$ should play a central role
for the $N=1$ supersymmetric extension of quantum
Teichm\"uller theory.

The plan of this paper is as follows. The first part is devoted to
$\Uqtwo$. The main purpose here is to review (and correct) the known
results and to explain how they are proved. This will help us later
to analyse representations of $\Uqosp$. In fact, with the right
notations, most formulas for the supersymmetric algebra resemble
those for $\Uqtwo$. In addition, the strategy of the proofs can
essentially be carried over form the non-supersymmetric theory.
The first part commences with a short description of $\Uqtwo$ and
its self-dual representations. Then we construct the Clebsch-Gordan
maps for the decomposition of products of self-dual representations
and prove their intertwining and orthonormality property. Section 4
contains the Racah-Wigner coefficients for $\Uqtwo$. The same steps
are then taken in the second part on the quantum deformed algebra
$\Uqosp$. Once more, we describe the algebra and a series of
self-dual representations whose Clebsch-Gordan coefficients are
determined in section 6. Our main result for the associated
Racah-Wigner coefficients is contained in section 7. Finally, we
compare our expressions for the Racah-Wigner coefficients of
$\Uqosp$ with the known formulas for the fusion matrix of $N=1$
Liouville field theory
\cite{Hadasz:2007wi,Chorazkiewicz:2008es,Chorazkiewicz:2011zd} and
find agreement. The paper concludes with a list of open problems
and further directions to explore.

\section{Self-dual continuous series for $\Uqtwo$}

The goal of this section is to introduce the quantum deformed enveloping algebra
$\Uqtwo$ along with a continuous series of representations that was first
discussed by Schmuedgen in \cite{Schmudgen:1995bx} and further analysed by
Ponsot and Teschner \cite{Ponsot:1999uf,Ponsot:2000mt}. It is self-dual under the
replacement $\ub \rightarrow \ub^{-1}$ in a sense to be made more precise below.

The q-deformed universal enveloping algebra $\Uqtwo$ of the Lie algebra
$sl(2)$ is generated by the elements $K, K^{-1}, E^\pm,$ with relations
\begin{align*}
 KE^\pm &= q^{\pm 1} E^\pm K,\\[2mm]
 [E^+,E^-] &= - \frac{K^2 - K^{-2}}{q-q^{-1}},
\end{align*}
where $q=e^{i\pi \ub^2}$ is the deformation parameter. We shall parametrize
the deformation through a real number $\ub$ so that $q$ takes values on the
unit circle. Given such a choice, the deformed algebra comes equipped with
the following *-structure
\begin{equation}
 K^* \ = \ K \quad , \quad
 (E^\pm)^* \ =\ E^\pm\ .
\end{equation}
The tensor product of any two representations can be built with the help of
the following co-product
\begin{equation}
 \Delta(K) \ = \ K\otimes K \quad , \quad
 \Delta(E^\pm) \ = \ E^\pm\otimes K + K^{-1}\otimes E^\pm\ .
\end{equation}
Finally, there is one more object we shall need below, namely the
quadratic Casimir element $C$ of $\mathcal{U}_q(sl(2))$ which reads
\begin{equation*}
C \ = \ E^-E^+ - \frac{q K^2 + q^{-1} K^{-2} +2}{(q-q^{-1})^2}\ .
\end{equation*}
Having collected the most important formulas concerning the algebraic
structure, we now want to introduce the series of representations we
are going to analyse in this work. It is parametrized by a label $\alpha$
that takes values in $\alpha \in \frac{Q}{2} + i \mathbb{R}$, where $Q$ is
related to the deformation parameter through $Q=\ub+\frac{1}{\ub}$. The carrier
spaces ${\mathcal P}_\alpha$ of the associated representations consist of
entire analytic functions $f(x)$ in one variable $x$ whose Fourier transform
$\hat f(\omega)$ is meromorphic in the complex plane with possible poles in
\begin{equation} \label{Salpha}
 {\mathcal S}_\alpha := \{\  \omega = \pm i(\a -Q -n\ub-m\ub^{-1});
n,m \in \mathbb{Z}^{\leq 0}\ \}\ .
\end{equation}
On this space, we represent the element $K$ through a shift
operator in the imaginary direction,
\begin{equation}
\pi_\alpha(K) \ = \ e^{\frac{i\ub}{2} \partial_x} \ =: \
T_x^{\frac{i \ub}{2}} \ .   \label{repK}
\end{equation}
By construction, the operator $T^{ia}_x$ defined in the previous
equation acts on functions $f \in {\mathcal P}_\alpha$ as
\begin{equation} \label{shiftT}
 T^{a}_x f(x) := f(x+a) \ .
 \end{equation}
The expressions for the remaining two generators $E^\pm$ are linear
combinations of two shift operators in opposite directions
\begin{equation} \label{repEpm}
 \pi_\alpha(E^\pm) \ = \ e^{\pm 2 \pi \ub x} \ \frac{e^{\pm i\pi \ub \alpha}\,
 T^\frac{i\ub}{2} - e^{\mp i\pi \ub \alpha}\, T^{-\frac{i\ub}{2}}}{q - q^{-1}}
 \ =: \ e^{\pm 2 \pi \ub x} [(2\pi)^{-1} \partial_x \pm \bar\alpha]_\ub \ .
\end{equation}
Here and in the following we shall use the symbol $\bar \alpha$ to
denote $\bar\alpha = Q-\alpha$ and we introduced the following
notation
\begin{equation} \label{[x]}
 \left[ x\right]_\ub \ = \
 \frac{\sin(\pi \ub x)}{\sin(\pi \ub^2)} .
\end{equation}
We claimed before that the representations $\pi_\alpha$ are self-dual
in a certain sense. Now we can make this statement more precise. To this
end, let us define a second action $\tilde \pi_\alpha$ of
${\mathcal{U}}_{\tilde q} (sl(2))$ with $\tilde q = \exp(i\pi/\ub^2)$ on the
space ${\mathcal P}_\alpha$ through the formulas \eqref{repK} and
\eqref{repEpm} with $\ub$ replaced by $\ub^{-1}$. Remarkably, the two actions
$\pi_\a$ and $\tilde \pi_{\alpha}$ commute with each other. This is the
self-duality property we were referring to.

\section{The Clebsch-Gordan coefficients for $\Uqtwo$}
The action $\pi_{\alpha_2}\otimes\pi_{\alpha_1}$ of the
quantum universal enveloping algebra $\Uqtwo$ on the tensor product of any
two representations $\pi_{\alpha_1}$ and $\pi_{\alpha_2}$ is defined in
terms of the coproduct, as usual. Such a tensor product is reducible and
its decomposition into a direct sum of irreducibles is what defines the
Clebsh-Gordan coefficients. In this case at hand, one has the following
decomposition,
\begin{equation*}
 {\mathcal P}_{\alpha_2} \otimes {\mathcal P}_{\alpha_1} \ \simeq \
 \int_{{\frac{Q}{2}} + i\mathbb{R^+}}^\otimes \ud \alpha_3\, {\mathcal P}_{\alpha_3} .
\end{equation*}
We are going to spell out and prove an explicit formula for the
homomorphism
\begin{equation*}
 f(x_2,x_1) \to F_f(\alpha_3, x_3) \ =\
 \int_\mathbb{R} \ud x_2 \ud x_1 \clebsh{\alpha_3}{x_3} f(x_2,x_1)\ .
\end{equation*}
Here, $f(x_2,x_1)$ denotes an element in ${\mathcal P}_{\alpha_2} \otimes
{\mathcal P}_{\a_1}$ and $F_f(\a_3,x_3)$ is its image in ${\mathcal P}_{\a_3}$.
In order state a formula for the Clebsch-Gordan map, we introduce
\begin{equation}\label{Ddef}
 D(\z;\a) = \frac{S_\text{b}(\z)}{S_\text{b}(z+\a)},
\end{equation}
and
\begin{align*}
 \z_{21} &= ix_{12} - Q +\frac{1}{2}(2\balpha_3 + \balpha_1 + \balpha_2),\\[1mm]
 \z_{31} &= ix_{13} + \frac{1}{2}(\balpha_1-\balpha_3),\\[1mm]
 \z_{32} &= ix_{32} + \frac{1}{2}(\balpha_2 - \balpha_3),
\end{align*}
where $\balpha_i \in Q\slash2 + i\mathbb{R}$ is defined as before through $\balpha_i
= Q-\alpha_i$ and we used $x_{ij} = x_i-x_j$. The symbols $\a_{ij}$ stand for
$$ \a_{21} = \a_1 + \a_2 + \a_3 - Q \quad , \quad
\a_{31} = Q + \a_1 - \a_2 - \a_3 \quad , \quad
\a_{32} = Q - \a_1 + \a_2 - \a_3 \ .  $$
With all these notations, we are finally able to spell out the relevant
Clebsh-Gordan coefficients \cite{Ponsot:2000mt},
\begin{equation}
  \clebsh{\alpha_3}{x_3}\ = \ \mathcal{N}
  D(\z_{21};\a_{21})D(\z_{23};\a_{23})D(\z_{13};\a_{13})\ ,
\end{equation}
where
\begin{equation} \label{Norm}
 \mathcal{N} \ = \ \exp\left[-\frac{i\pi}{2}(\balpha_3\alpha_3 - \balpha_2\alpha_2-\balpha_1\alpha_1)\right]\ .
\end{equation}
Let us note that this product form of the Clebsch-Gordan coefficients is
familiar e.g. from the 3-point functions in conformal field theory which
may be written as a product. Although the representations we study here
are not obtained by deforming discrete series representations of $sl(2)$,
i.e.\ of those representations that fields of a conformal field theory
transform in, the familiar product structure of the Clebsch-Gordan
coefficients survives.

\subsection{The intertwining property}

The fundamental intertwining property of the Clebsch-Gordan coefficients
takes the following form
\begin{equation} \label{int}
 \pi_{\alpha_3}(X) \clebsh{\alpha_3}{x_3} =
  \clebsh{\alpha_3}{x_3}(\pi_{\alpha_2}\otimes\pi_{\alpha_1})\Delta(X)
\end{equation}
for $X = K, E^\pm$. The equation should be interpreted as an identity of
operators in the representation space ${\mathcal P}_{\a_2} \otimes
{\mathcal P}_{\a_1}$. While the operators $K$ and $E^\pm$ may be
expressed through multiplication and shift operators, the Clebsch-Gordan
map itself provides the kernel of an integral transform. With the help of
partial integration, we can re-write the intertwining relation as an
identity for the integral kernel,
\begin{equation} \label{int2}
 \pi_{\alpha_3}(X) \clebsh{\alpha_3}{x_3} = (\pi_{\alpha_2}\otimes
 \pi_{\alpha_1})\Delta^t(X) \clebsh{\alpha_3}{x_3},
\end{equation}
where the superscript $^t$ means that we should replace all shift
operators by shifts in the opposite direction, i.e. $(T_x^{ia})^t
=T_x^{-ia}$ and exchange the order between multiplication and
shifts, i.e $(f(x)T_x^{ia})^t = T_x^{-ia} f(x)$. In this new
form, the intertwining property is simply an identity of functions
in the variables $x_i$.

One can check eq.\ \eqref{int2}  by direct computation. This is
particularly easy for the element $K$ for which eq. \eqref{int2}
reads
\begin{equation}
 T^{\frac{i \ub}{2}}_{x_3} \clebsh{\alpha_3}{x_3} \ = \
 T^{-\frac{i \ub}{2}}_{x_2} T^{-\frac{i \ub}{2}}_{x_1}
 \clebsh{\alpha_3}{x_3} \ .
\end{equation}
Since the Clebsch-Gordan maps depend only in the differences $x_{ij}$
we can replace $T_{x_1} = T_{12} T_{13}$ etc.\ where $T_{ij}$ denotes
a shift operator acting on $x_{ij}$. Consequently, the intertwining
property for $K$ becomes
\begin{equation}
 T^{-\frac{i \ub}{2}}_{13} T^{-\frac{i \ub}{2}}_{23} \clebsh{\alpha_3}{x_3} \ = \
 T^{\frac{i \ub}{2}}_{12} T^{-\frac{i \ub}{2}}_{23} T^{-\frac{i \ub}{2}}_{12}
 T^{-\frac{i \ub}{2}}_{13} \clebsh{\alpha_3}{x_3},
\end{equation}
which is trivially satisfied since all shifts commute. This concludes
the proof of the intertwining property \eqref{int2} for $X=K$.

For $X=E^+$ the check is a bit more elaborate. Using the anti-symmetry
$[-x]_\ub = - [x]_\ub$ of the function \eqref{[x]} and the property
$\partial_x^t = - \partial_x$ of derivatives, we obtain
\begin{align*}
& e^{2\pi \ub x_3} [\delta_{x_3} + \balpha_3]_\ub \clebsh{\alpha_3}{x_3} =\\[2mm]
 &  - [\delta_{x_2} - \balpha_2]_\ub \, e^{2\pi b x_2} \, T^{\frac{i\ub}{2}}_{x_1}
  \clebsh{\alpha_3}{x_3} -  [\delta_{x_1} - \balpha_1]_\ub \, e^{2\pi \ub x_1}
  \, T^{-\frac{i\ub}{2}}_{x_2} \clebsh{\alpha_3}{x_3}\ .
\end{align*}
where $\delta_x = (2\pi)^{-1} \partial_x$. After a bit of rewriting we
find
\begin{align*}
  &\left[\, e^{i\pi \ub(\balpha_1-\balpha_2)\slash2} [-ix_{21} + Q -\frac{1}{2}(\balpha_2 + \balpha_1)]_\ub T^{i\ub}_{21} T^{ib}_{23} \right.\\[2mm]  &\,  \ + e^{-\pi \ub x_{23}} e^{-i\pi \ub (\balpha_3 + \balpha_1)\slash2} [-ix_{13} + Q + \frac{1}{2}(\balpha_3-\balpha_1)]_\ub T^{i\ub}_{13} T^{i\ub}_{23}  \\[2mm]
 &\left. \, \ - e^{-i\pi \ub Q} e^{-\pi \ub x_{13}} e^{i\pi \ub (\balpha_2 + \balpha_3)\slash2} [-ix_{23} + \frac{1}{2}(\balpha_2 - \balpha_3)]_\ub \right] \clebsh{\alpha_3}{x_3} = 0.
\end{align*}
Now, because of the shift properties of the function $S_\ub$, see Appendix A.1, we have
\begin{equation*}
 T^{i\ub}_x \frac{S_\ub(-ix+a_1)}{S_\ub(-ix+a_2)} = \frac{[-ix + a_1]_\ub}{[-ix + a_2]_\ub} \frac{S_\ub(-ix+a_1)}{S_\ub(-ix+a_2)} T^{i\ub}_x .
\end{equation*}
With the help of this equation it is easy to check that our Clebsch-Gordan coefficients
obey the desired intertwining relation with $E^+$. For the intertwining property involving
$X=E^-$ one proceeds in a similar way.

\subsection{Orthogonality and Completeness}

The Clebsh-Gordan coefficients for the self-dual series of $\Uqtwo$ satisfy the
following orthogonality and completeness relation
\begin{equation}
\int_\mathbb{R}\ud x_2 \ud x_1 \left[^{\alpha_3}_{x_3} \phantom{}^{\alpha_2}_{x_2} \phantom{}^{\alpha_1}_{x_1} \right]^* \left[^{\beta_3}_{y_3} \phantom{}^{\alpha_2}_{x_2} \phantom{}^{\alpha_1}_{x_1} \right]
= |S_\ub(2\alpha_3)|^{-2}\delta(\alpha_3 - \beta_3)\delta(x_3 - y_3), \label{ortho}
\end{equation}
\begin{equation}
 \int_{{\frac{\mathbb{Q}}{2}} + i\mathbb{R^+}} \ud\alpha_3 \int_\mathbb{R}\ud x_3 |S_\ub(2\alpha_3)|^2 \left[^{\alpha_3}_{x_3} \phantom{}^{\alpha_2}_{x_2} \phantom{}^{\alpha_1}_{x_1} \right]^* \left[^{\alpha_3}_{x_3} \phantom{}^{\alpha_2}_{y_2} \phantom{}^{\alpha_1}_{y_1} \right] = \delta(x_2-y_2) \delta(x_1-y_1)\ .  \label{complete}
\end{equation}
Except for the normalizing factor on the right hand side of eq.\ \eqref{ortho},
these relations follow from the intertwining properties of Clebsch-Gordan maps.
Since the complete proof of eqs.\ \eqref{ortho} and
\eqref{complete} has not been spelled out in the literature we will discuss
the derivation of eq.\ \eqref{ortho} in some detail here. This will allow us
to skip over some details later when we discuss the corresponding issues for
the deformed superalgebra. In order to compute the integral on the left hand
side of eq.\ \eqref{ortho} we shall employ a star-triangle relation for the
functions $S_\ub$ along with several of its corollaries. All necessary
integral formulas are collected in Appendix B.1.

Before we proceed proving the orthogonality relations, let us point out that
the equations \eqref{ortho} involve products of the Clebsch-Gordan kernels.
Since these are distributional kernels, one must take some care when
multiplying two of them. Following \cite{Ponsot:2000mt}, the strategy is to
regularize the Clebsch-Gordan maps through some $\epsilon$ prescription, then
to multiply the regularized kernels before we send the parameter $\epsilon$ to
zero in the very end of the computation. For the problem at hand, one
appropriate regularization takes the form
\begin{equation} \label{regCG}
  \clebsh{\alpha_3}{x_3}_\epsilon \ = \ \mathcal{N} D(\z_{21}+ \epsilon;\a_{21}-\epsilon) D(\z_{32}+\epsilon;\a_{32}-\epsilon)D(\z_{31};\a_{31}+\epsilon),
\end{equation}
with the same normalization \eqref{Norm} as above. Our prescription is different from the
one used in \cite{Ponsot:2000mt}.

Inserting the regularized Clebsch-Gordan maps into the orthogonality
relation \eqref{ortho}, we obtain
\begin{align*}
 &\int \ud x_2\ud x_1 \clebsh{\alpha_3}{x_3}^*_\epsilon \clebsh{\beta_3}{y_3}_\epsilon =
  \eta \int\ud x_2 \ud x_1
 \frac{S_\ub(-ix_{21} + Q -\frac{1}{2}(\balpha_2 + \balpha_1))}
 {S_\ub(-ix_{21} + Q -\frac{1}{2}(\balpha_2 + \balpha_1))} \ \times\\[2mm]
 &\times S_\ub(-ix_{21}- Q +\frac{1}{2}(2\bbeta_3 + \balpha_1 + \balpha_2) + \epsilon)
 \ S_\ub(ix_{21} + Q -\frac{1}{2}(2\balpha_3 + \balpha_1 + \balpha_2) + \epsilon)\, \times \\[2mm]
 & \times S_\ub(-i(x_{1}-y_3) + Q +\frac{1}{2}(\balpha_1-\bbeta_3-2\balpha_2) - \epsilon)
 S_\ub(ix_{13} -\frac{1}{2}(\balpha_1-\balpha_3-2\balpha_2)-\epsilon)\times \\[2mm]
&\times S_\ub(-ix_{13} + \frac{1}{2}(\balpha_3-\balpha_1))\ S_\ub(-i(x_{1}-y_3) - \frac{1}{2}(\bbeta_3-\balpha_1))\ \times \\[2mm]
& \times \ D^\ast(z_{32}+\epsilon,\a_{32}-\epsilon) \
D(\tilde z_{32} +\epsilon, \tilde \a_{32}-\epsilon) \ =: \ I^\epsilon_1 \ .
\end{align*}
In writing this expression we have expressed all the D-functions that contain
some dependence on the variable $x_1$ in terms of $S_\ub$, see eq.\ \eqref{Ddef}.
We brought all but three of the $S_\ub$ functions to the numerator with the help of
the property $S^{-1}_\ub(x) = S_\ub(Q-x)$. In taking the complex conjugate, we
used that the variables $x_i,y_3$ and our regulator $\epsilon$ are real. The
labels $\alpha_i$ and $\beta_3$, on the other hand, satisfy $\alpha_i^\ast
= \bar \alpha_i = Q-\alpha_i$ and $\beta_3^\ast = \bar \beta_3 = Q-\beta_3$.
Finally, we introduced $\tilde {\mathcal N}, \tilde z_{32}$ and $\tilde
\alpha_{32}$. These are obtained from ${\mathcal N}, z_{32}$ and $\alpha_{32}$
by the substitution $x_3 \rightarrow y_3$ and $\a_3 \rightarrow \beta_3$. The
constant prefactor $\eta$ is given by $\eta = {\mathcal N}\tilde{\mathcal N}$.

Before we continue our evaluation of the integrals we note that the fraction
of $S$-functions in the first line of the previous equation cancels out. Hence,
we are left with a product of six $S$-functions that contain all the $x_1$
dependence of the integrand. It turns out that we can actually evaluate the
$x_1$ integral with the help of the following star-triangle equation, see
e.g.\ \cite{kashaev},
\begin{equation} \label{startriangle}
 \int \ud x_1 \prod_{i=1}^3 S_\ub(ix_1+\gamma_i)S_\ub(-ix_1+\delta_i)
 = \prod_{i,j=1}^3 S_\ub(\gamma_i+\delta_j)
\end{equation}
which holds as long as the arguments on the left hand side add up to
$Q$, i.e.\ if
\begin{equation*}
 \sum^3_{i=1}(\gamma_i+\delta_i) = Q,
\end{equation*}
It is not difficult to check that the arguments which appear in our formula
for $I^\epsilon_1$ above satisfy this condition. Hence, we can perform the integral
over $x_1$ to obtain
\begin{align*}
  I^\epsilon_1 &= \eta S_\ub(\bbeta_3-\balpha_3 + 2\epsilon)
  \frac{S_\ub(-i(y_3-x_3)+\frac{1}{2}(\balpha_3-\bbeta_3))}
  {S_\ub(-i(y_3-x_3)-\frac{1}{2}(\balpha_3-\bbeta_3) +2\epsilon)} \frac{S_\ub(\balpha_3+\balpha_2-\balpha_1 - \epsilon)}{S_\ub(\bbeta_3+\balpha_2-\balpha_1+\epsilon)} \ \times \\[2mm]
&\times \int \ud x_2  \frac{S_\ub(-ix_{23} - Q + \frac{1}{2}(2\bbeta_3+\balpha_2+\balpha_3)+\epsilon)}{S_\ub(-i(x_{2}-y_3) + \frac{1}{2}(2\balpha_3+\balpha_2+\bbeta_3)-\epsilon))} \frac{S_\ub(-i(x_{2}-y_3) + \frac{1}{2}(\balpha_2 - \bbeta_3)+ \epsilon)}{S_\ub(-ix_{23} +Q + \frac{1}{2}(\balpha_2 - \balpha_3)-\epsilon)} = \\[2mm]
&= \eta S_\ub(\bbeta_3-\balpha_3 + 2\epsilon) \frac{S_\ub(-i(y_3-x_3)+\frac{1}{2}(\balpha_3-\bbeta_3) )}{S_\ub(-i(y_3-x_3)-\frac{1}{2}(\balpha_3-\bbeta_3) +2\epsilon)} \frac{S_\ub(\balpha_3+\balpha_2-\balpha_1-\epsilon)}{S_\ub(\bbeta_3+\balpha_2-\balpha_1
+\epsilon)} \ \times \\[2mm]
&\times \int \frac{\ud \tau}{i} \frac{S_\ub(\tau + \xi_1 + \epsilon)}{S_\ub(Q+\tau+\xi_2-
\epsilon)} \frac{S_\ub(\tau-\xi_1+\epsilon)}{S_\ub(Q+\tau-\xi_2-\epsilon)} =: I^\epsilon_2\ .
\end{align*}
In the first step we evaluated the right hand side of the star-triangle relation
\eqref{startriangle} and we expressed the remaining two $D$-functions that appear
in $I^\epsilon_1$ through the functions $S_\ub$. After these two steps, the formula for
$I^\epsilon_1$ should contain a total number of $9+4 = 13$ functions $S_\ub$. It turns out
that four of them cancel against each other so that we are left with the nine
factors in the first two lines of the previous formula. In passing to the lower
lines we simply performed the substitutions
\begin{align*}
 \tau &= -i x_2 + \balpha_2\slash2 -i(x_3+y_3)\slash2 -Q\slash2 +
 (\bbeta_3+\balpha_3)\slash4,\\[2mm]
 \xi_1 &= -\frac{i}{2}(y_3-x_3) + \frac{1}{4}(3\bbeta_3+\balpha_3)
 - \frac{Q}{2},\\[2mm]
 \xi_2 &= \frac{i}{2}(y_3-x_3) + \frac{1}{4}(\bbeta_3+3\balpha_3) - \frac{Q}{2}.
\end{align*}
In this form we can now also carry out the integral of the variable
$\tau$ using a limiting case of the Saalsch\"utz formula, see Appendix B.1,
to find
\begin{align*}
   I^\epsilon_2 & = \eta S_\ub(\gamma + 2\epsilon) \frac{S_\ub(-\xi_- - \gamma )}
 {S_\ub(2\epsilon -\xi_-)} \frac{S_\ub(\balpha_3+\balpha_2-\balpha_1 - \epsilon)}
   {S_\ub(\gamma +\balpha_3 + \balpha_2
   -\balpha_1 +\epsilon)} \ \times \\[2mm]
&\times e^{-i\pi \xi_-\xi_+}
\frac{S_\ub(2\epsilon - \xi_-) S_\ub(2\epsilon + \xi_-) S_\ub(2\epsilon - \xi_+)
S_\ub(2\epsilon + \xi_+)}{S_\ub(4\epsilon)}
\end{align*}
where $\gamma = \bbeta_3-\balpha_3 \in i \mathbb{R}$ and
\begin{align*}
 \xi_-&=\xi_2-\xi_1= i(y_3-x_3) - \frac{1}{2} \, \gamma \in i \mathbb{R} ,\\[2mm]
 \xi_+&=\xi_2+\xi_1= \bbeta_3+\balpha_3-Q \in i \mathbb{R}\setminus \{0\} .
\end{align*}
Having performed both integrations, it remains to remove our
regulator $\epsilon$. The most nontrivial part of this computation
is to show that
\begin{equation}
 \lim_{\epsilon\rightarrow0} \frac{S_\ub(2\epsilon+\gamma)
 S_\ub( -\xi_- -\gamma) S_\ub(2\epsilon + \xi_-) }{S_\ub(4\epsilon)}
 = \delta(i\gamma)\delta(i\xi_-)\ .
\end{equation}
A full proof is given in Appendix C. The remaining factors in $I^\epsilon_2$
possess a regular limit. In particular we find
\begin{equation}
 \lim_{\epsilon\rightarrow0} S_\ub(2\epsilon - \xi_+) S_\ub(2\epsilon + \xi_+)
  =\frac{1}{S_\ub(Q+\xi_+)S_\ub(Q-\xi_+)} =
 |S_\ub(\bbeta_3+\balpha_3)|^{-2} .
\end{equation}
Finally, for the normalization factor $\eta = {\mathcal N}\tilde{ \mathcal N}$ we
obtain
\begin{align*}
 \eta & = e^{-\frac{i\pi}{2}(\balpha_3\alpha_3 - \balpha_2\alpha_2-\balpha_1\alpha_1)}
 e^{\frac{i\pi}{2}(\bbeta_3\beta_3 - \balpha_2\alpha_2-\balpha_1\alpha_1)} =
 e^{-\frac{i\pi}{2}\gamma(\gamma+2\balpha_3-Q)},
\end{align*}
Putting all these results together we have shown that
\begin{align*}
 &\lim_{\epsilon\rightarrow0} \int \ud x_2 \ud x_1 \clebsh{\alpha_3}{x_3}^*_\epsilon
 \clebsh{\beta_3}{y_3}_\epsilon = \\[2mm] & \hspace*{2cm} = e^{-i\pi \xi_-\xi_+} \frac{e^{-\frac{i\pi}{2}\gamma(\gamma+2\balpha_3-Q)}}
 {|S_\ub(\bbeta_3+ \balpha_3)|^{2}} \frac{S_\ub(\balpha_3+\balpha_2-\balpha_1)}
 {S_\ub(\gamma+\balpha_3+\balpha_2-\balpha_1)}  \delta(i\gamma)\delta(i\xi_-)\\[2mm]
 & \hspace*{4.8cm} = |S_\ub(\bbeta_3+ \balpha_3)|^{-2} \delta(i (\bbeta_3-\balpha_3)
 \delta(y_3-x_3)\ .
\end{align*}
This is the orthonormality relation we set out to prove. The proof of eq.\
\eqref{complete} is left as an exercise, see \cite{Ponsot:2000mt} for some
helpful comments. An alternative proof of the orthonormality relation was
published recently in \cite{Nidaiev:2013bda}, \cite{Derkachov:2013cqa}.

\section{The Racah-Wigner coefficients for $\Uqtwo$}

The Racah-Wigner coefficients describe a change of basis in the 3-fold
tensor product of representations. Let us denote these three representations
by $\pi_{\alpha_i}, i=1,2,3$. In decomposing their product into irreducibles
$\pi_{\alpha_4}$ there exists two possible fusion paths, denoted by $t$ and
$s$, which are described by the following combination of Clebsch-Gordan
coefficients
\begin{align}
\label{Phit}
\Phi^t_{\alpha_t} \left[{\alpha_3 \atop \alpha_4}
{\alpha_2\atop\alpha_1}\right]_\epsilon(x_4; x_i)
& = \int \ud x_t \left[ {\alpha_4 \atop x_4}{\alpha_t \atop x_t}
{\alpha_1 \atop x_1}\right]_{\epsilon}
\left[ {\alpha_t \atop x_t}{\alpha_3 \atop x_3}
{\alpha_2 \atop x_2}\right]_{\epsilon},\\[2mm]
\label{Phis}
 \Phi^s_{\alpha_s}
 \left[{\alpha_3 \atop \alpha_4}{\alpha_2\atop\alpha_1}\right]_\epsilon(x_4; x_i)
 & = \int \ud x_s \left[ {\alpha_4 \atop x_4}{\alpha_3 \atop x_3}
 {\alpha_s \atop x_s}\right]_{\epsilon}
 \left[ {\alpha_s \atop x_s}{\alpha_2 \atop x_2}
 {\alpha_1 \atop x_1}\right]_{\epsilon}.
\end{align}
The regularization we use here is the same as in the previous
section. From the two objects $\Phi^s$ and $\Phi^t$ we obtain
the Racah-Wigner coefficients as
\begin{equation} \label{6jint}
 \left\{ {\alpha_1 \atop \alpha_2} {\alpha_3\atop\alpha_4} |
 {\alpha_s \atop \alpha_t}\right\}_\ub =
 \lim_{\epsilon\to0} \int \ud\alpha'_4\ud x'_4
 \int \ud^3 x_i \Phi^t_{\alpha_t} \left[{\alpha_3 \atop \alpha'_4}{\alpha_2\atop\alpha_1}\right]^*_\epsilon\!\!(x'_4; x_i) \Phi^s_{\alpha_s} \left[{\alpha_3 \atop \alpha_4}{\alpha_2\atop\alpha_1}\right]_\epsilon\!\!(x_4; x_i) .
\end{equation}
After inserting the concrete expressions \eqref{regCG} for the regularised
Clebsch-Gordan coefficients one may evaluate the integrals to obtain
\cite{Ponsot:2000mt}
\begin{eqnarray}\nonumber
 \left\{ {\alpha_1 \atop \alpha_2} {\alpha_3\atop\alpha_4}
 | {\alpha_s \atop \alpha_t}\right\}_\ub
  & = &  |S_\ub(2\alpha_4)|^2
   \frac{S_\ub(a_4) S_\ub(a_1)}{S_\ub(a_2)S_\ub(a_3)}\ \times \\[2mm]
   & & \quad \times \, \int_{i\mathbb{R}} \ud t
  \frac{S_\ub(u_4+t)S_\ub(\tilde u_4+t)S_\ub( u_3+t)S_\ub(\tilde u_3+t)}
  {S_\ub(u_{23} +t)S_\ub(\tilde u_{23}+t)
  S_\ub(2\alpha_s+t)S_\ub(Q+t)} \label{RWsltwo}
\end{eqnarray}
where the four variables $a_i$ are associated with the four Clebsch-Gordan
maps that appear in eqs.\ \eqref{Phit} and \eqref{Phis}
\begin{align*}
a_1 & =\alpha_1-\bar\alpha_t+\alpha_4 \quad ,\quad
& a_2  =\alpha_2-\alpha_3+\alpha_t \\[2mm]
a_3 & =\alpha_3 - \bar\alpha_s + \alpha_4\quad , \quad
& a_4  = \alpha_2 - \alpha_1 + \alpha_s
\end{align*}
and similarly for the remaining set of variables,
\begin{align}
 & u_4  = \alpha_s+\alpha_1-\alpha_2\quad ,
 & \tilde u_4 = \alpha_s+\bar\alpha_1-\alpha_2,\nonumber \\[2mm]
 & u_3  = \alpha_s+\alpha_4-\alpha_3 \quad ,
& \tilde u_3 = \alpha_s+\alpha_4 -\bar \alpha_3, \\[2mm]
& u_{23} = \alpha_s+\alpha_t+\alpha_4-\alpha_2\quad ,
& \tilde u_{23} = \alpha_s+\bar\alpha_t+\alpha_4-\alpha_2\ .
\nonumber
\end{align}
Note that the first factor in our formula \eqref{RWsltwo} for the
Racah-Wigner symbols differs from \cite{Ponsot:2000mt}. Our expression
is a result of carefully analyzing the integrals the definition
\eqref{6jint} of the Racah-Wigner symbols. The same normalization
was found independently by Nidaiev in \cite{nidaiev}.

The derivation of eq.\ \eqref{RWsltwo} from eq.\ \eqref{6jint} is
in principle straightforward, though a bit cumbersome. One simply
has to evaluate the integrals. The integrals over the variables
$x_i, i=1,2,3,$ are performed with the help of Cauchy's integral
formula. The resulting integral expression involves delta functions
in both the difference $\a_4-\a_4'$ and $x_4-x_4'$. Hence the
integrals over $x'_4$ and $\a'_4$ are easy to perform at the
end of the computation. So, let us get back to the integrals
over $x_i, i=1,2,3$. It is convenient so start with $x_1$. In
order to perform the integration, one needs to keep track of
all the poles in the integrand along with their residues. Since
the functions $\Phi^s$ and $\Phi^t$ are ultimately built from
$S_\ub$ through equations \eqref{Ddef}, \eqref{regCG},
\eqref{Phit} and \eqref{Phis}, this step only requires
knowledge of the poles and residues of $S_\ub$. All this
information on $S_\ub$ can be found in Appendix A.1. Once the
integration over $x_1$ has been performed, one focuses on the
variable $x_3$. There are a few poles that have been around
before we integrated over $x_1$. In addition,
the integration over $x_1$ brought in some new poles through
the usual pole collisions (pinching). These must all be
accounted for before we can apply Cauchy's formula again to
perform the integration over $x_3$. Similar comments apply
to the final integral over $x_2$. Many more details on this
computations can be found in section 5 of the paper by Ponsot
and Teschner \cite{Ponsot:2000mt}. Let us stress again that no
fancy identities are needed at any stage of the calculation.

After these comments on the derivation of eq.\ \eqref{RWsltwo}, let us list a
few more properties of the Racah-Wigner symbols. To begin with, they can be
shown to satisfy the following orthogonality relations
\begin{equation*}
 \int_{{\frac{\mathbb{Q}}{2}} + i\mathbb{R^+}} \ud\alpha_s | S_\ub(2\alpha_s)|^2 \left( \left\{ {\alpha_1 \atop \alpha_3} {\alpha_2\atop\alpha_4} | {\alpha_s \atop \beta_t}\right\}_\ub \right)^* \left\{ {\alpha_1 \atop \alpha_3} {\alpha_2\atop\alpha_4} | {\alpha_s \atop \alpha_t}\right\}_\ub = |S_\ub(2\alpha_t)|^2 \delta(\alpha_t-\beta_t).
\end{equation*}
As a consequence of their very definition, the Racah-Wigner symbols must
also satisfy the pentagon equation
\begin{equation*}
 \int_{{\frac{\mathbb{Q}}{2}} + i\mathbb{R^+}} \ud \delta_1 \left\{ {\alpha_1 \atop \alpha_3} {\alpha_2\atop\alpha_4} | {\beta_1 \atop \delta_1}\right\}_\ub \left\{ {\alpha_1 \atop \alpha_4} {\delta_1\atop\alpha_5} | {\beta_2 \atop \gamma_2}\right\}_\ub \left\{ {\alpha_2 \atop \alpha_4} {\alpha_3\atop\gamma_2} | {\delta_1 \atop \gamma_1}\right\}_\ub = \left\{ {\beta_1 \atop \alpha_4} {\alpha_3\atop\alpha_5} | {\beta_2 \atop \gamma_1}\right\}_\ub \left\{ {\alpha_1 \atop \gamma_1} {\alpha_2\atop\alpha_5} | {\beta_1 \atop \gamma_2}\right\}_\ub\ .
\end{equation*}
More recently, Teschner and Vartanov found an interesting alternative
expression for the Racah-Wigner coefficients \cite{Teschner:2012em}.
We will discuss this representation along with its extension to the
supersymmetric case in an accompanying paper.

\section{Self-dual continuous series for $\Uqosp$}

After this extended review of the quantum deformed enveloping algebra $\Uqtwo$
and its self-conjugate series of representations we are now well prepared to
turn to the algebra $\Uqosp$. We shall restate its definition here before
we describe a 1-parameter series of self-conjugate representations.

Following \cite{Kulish:1988gr}, the quantum deformed superalgebra $\Uqosp$ is
generated by the bosonic generators $K, K^{-1}$ along with two fermionic (odd) ones
$v^{(\pm)}$. These satisfy the relations
\begin{align*}
 K v^{(\pm)} &= q^{\pm1} v^{(\pm)} K,\\[1mm]
 \{ \vp, \vm \} &= -\frac{K^2 - K^{-2}}{q^{1\slash2} - q^{-1\slash2}},
\end{align*}
where $q=e^{i\pi \ub^2}$, as before. The similarity with the defining relations
of $\Uqtwo$ is striking, except that the elements $v^{\pm }$ are fermionic (odd)
so that we prescribe the anti-commutator $\{.,.\}$ of $\vp$ with $\vm$ instead of
the commutator. The algebraic relations are compatible with the following star
structure
\begin{equation}
  K^* \ = \ K\quad , \quad   v^{(\pm) *} \ = \ v^{(\pm)},
\end{equation}
and with the coproduct
\begin{equation}
 \Delta(K) \  = \ K\otimes K \quad ,\quad
 \Delta(v^{(\pm)}) \ = \ v^{(\pm)}\otimes K + K^{-1}\otimes v^{(\pm)},
\end{equation}
that can be used to define tensor products of representations. It is easy to
verify that the following even element of $\Uqosp$ commutes with all generators,
\begin{equation*}
  C = -\frac{q K^4 + q^{-1} K^{-4} + 2}{(q-q^{-1})^2} + \frac{(q K^2 + q^{-1} K^{-2}) v^{(-)} v^{(+)}}{q^{\frac{1}{2}}+q^{-\frac{1}{2}}} + v^{(-) 2} v^{(+) 2},
\end{equation*}
i.e.\ $C$ is a Casimir element. In addition, the algebra $\Uqosp$ also contains
an element $Q$ which is defined as
\begin{equation*}
 Q = \frac{1}{2}(\vm\vp - \vp\vm) + \frac{1}{2} \frac{K^2 + K^{-2}}{q^{1\slash2}
 + q^{-1\slash2}} .
\end{equation*}
Up to some shift, the element $Q$ may be considered as the square root of the
quadratic Casimir element $C$,
\begin{equation*}
 C = -\left(Q+\frac{2i}{q-q^{-1}}\right)\left(Q-\frac{2i}{q-q^{-1}}\right).
\end{equation*}
This concludes our short description of the algebraic setup so that we can
begin to discuss the representations we are about to analyse. Following the
intuition developed from the non-supersymmetric case, we shall introduce
representation on carrier spaces $\mathcal{Q}_\alpha$ which are parametrized
by a single parameter $\a$ of the form $\alpha \in \frac{Q}{2} +i\mathbb{R}$
where $Q=\ub+\frac{1}{\ub}$ and the relation between $\ub$ and $q$ is the same as
before. The spaces $\mathcal{Q}_\a$ are graded vector spaces. By definition,
they consist of pairs $(f^0(x),f^1(x))$ of functions $f^j$ which are
entire analytic and whose Fourier transform $\hat f^j(\omega)$ is allowed to
possess poles in the set $\mathcal{S}_\alpha$ that was defined in eq.\
\eqref{Salpha}. The upper index $j$ indicates the parity of the
element, i.e.\ vectors of the form $(f^0,0)$ are considered even while
we declare elements of the form $(0,f^1)$ to be odd. On these carrier spaces,
we represent the elements $K$ and $v^{\pm}$ through
\begin{equation} \label{rep_supersym}
 \pi_\alpha\left(K\right) = T_x^{\frac{i \ub}{2}}
 \left( \begin{array}{cc} 1 & 0 \\ 0 & 1 \end{array} \right)\quad ,\quad
 \pi_\alpha\left(v^{(\pm)}\right)  = i e^{\pm\pi \ub x}
 \left( \begin{array}{cc} 0 & [\delta_x \pm \bar\alpha]_- \\[2mm]
 \left[\delta_x \pm \bar\alpha\right]_+ & 0 \end{array} \right)
\end{equation}
where $T^{ia}_x$ denotes the shift operator that was defined in eq.\
\eqref{shiftT} and we introduced
\begin{equation}
 \left[ x\right]_- \ = \ \frac{\sin(\frac{\pi \ub x}{2})}
 {\sin(\frac{\pi \ub^2}{2})}\quad ,\quad
 [x]_+  = \frac{\cos(\frac{\pi \ub x}{2})}
 {\cos(\frac{\pi \ub^2}{2})} .
\end{equation}
Consequently, the matrix elements in our expression for $\pi_\a(v^\pm)$
are given by
\begin{align*}
 [\delta_x + \bar\alpha]_- &= \frac{e^\frac{i\pi \ub \a}{2}
 T^\frac{i\ub}{2} - e^\frac{-i\pi \ub \a}{2} T^{-\frac{i\ub}{2}}}
 {q^\frac{1}{2} - q^{-\frac{1}{2}}}, \\[2mm]
 [\delta_x + \bar\alpha]_+ &= \frac{e^\frac{i\pi \ub \a}{2}
 T^\frac{i\ub}{2} + e^\frac{-i\pi \ub \a}{2}
 T^{-\frac{i\ub}{2}}}{q^\frac{1}{2} + q^{-\frac{1}{2}}} .
\end{align*}
It is not difficult to check
that our prescription respects the algebraic relations in the universal
enveloping algebra $\Uqosp$ and
hence provides a family of representations. In these representations, we can
evaluate the Casimir element $C$ and its square root $Q$,
\begin{equation*}
 \pi_\alpha(C) = [\frac{Q}{2} - \alpha]^2_- [\frac{Q}{2} - \alpha]^2_+
 \sigma_0 \quad , \quad
  \pi_\alpha(Q) = [\bar\alpha - \frac{\ub}{2}]_- [\bar\alpha - \frac{\ub}{2}]_+
  \sigma_3,
\end{equation*}
where $\sigma_0$ denotes the 2-dimensional identity matrix and $\sigma_3$ is
the Pauli matrix that is diagonal in our basis. Note that the
value of the Casimir element $C$ is the same in the representations $\pi_\a$ and
$\pi_{\balpha}$. This is because the representations are actually equivalent.
In fact, one may easily check that the following unitary operator
\begin{equation*}
 \mathcal{I}_\alpha = \left( \begin{array}{cc} 0 &
 \frac{S_1(\alpha - i \omega)}{S_1(\balpha - i \omega)} \\
 \frac{S_0(\alpha - i \omega)}{S_0(\balpha - i \omega)} & 0 \end{array} \right),
\end{equation*}
involving the special functions $S_\nu$ defined in Appendix A.2, satisfies
\begin{equation*}
 \pi_{\balpha}(X)\mathcal{I}_\alpha = \mathcal{I}_\alpha\pi_{\alpha}(X)\quad ,
 \quad \mbox{for}\quad  X= K, v^{(\pm)}.
\end{equation*}
In order to discuss the reality properties of our representation, we need to
introduce the following matrix
\begin{equation}
 \lambda^2 =  \left( \begin{array}{cc}\varrho & 0 \\ 0 & \varrho^{-1}
 \end{array} \right)
\quad \mbox{ where } \quad \varrho = i\frac{q^{1/2}-q^{-1/2}}
{q^{1/2}+q^{-1/2}}\ .
\end{equation}
A square root $\lambda$ of the matrix $\lambda^2$ appears in the definition
of the scalar product
\begin{equation}\label{eq:sp}
 \langle g, f \rangle = \sum_{i,j} \int \ud x {g^i(x)}^\ast
 \lambda_{ij}  f^j(x)\ .
\end{equation}
A short calculation shows that the adjoint with respect to this scalar product
implements the $\ast$ operation we defined above, i.e. $\langle g, X f \rangle
= \langle X g, f\rangle$ for all three generators of $\Uqosp$. Once again one
can check that the representations $\pi_\alpha$ admit duality $\ub\to \ub^{-1}$ in
the same sense as above. More concretely, our formulas can be used to define a
representation of $\mathcal{U}_{\tilde q}(osp(1|2))$ with $\tilde q= e^{i\pi
\ub^{-2}}$ on $\mathcal{Q}_\alpha$ such that the corresponding operators
(anti-)commute with the representation operators for the original action
of $\Uqosp$ on  $\mathcal{Q}_\a$. Let us mention that a similar series of
representations was recently discussed in \cite{Ip:2013qba}, though the
precise relation to the ones we consider here not clear to us.

\section{The Clebsch-Gordan coefficients for $\Uqosp$}

As in the case of $\Uqtwo$, we are interested in the Clebsch-Gordan
decomposition of the representation $\pi_{\alpha_2}\otimes
\pi_{\alpha_1}$,
\begin{equation*}
 \pi_{\alpha_2} \otimes \pi_{\alpha_1} \simeq
 \int_{{\frac{\mathbb{Q}}{2}} + i\mathbb{R^+}}^\otimes \ud \alpha_3 \,
 \pi_{\alpha_3}\ .
\end{equation*}
We shall show below that there exist two independent intertwiners for
any given choice of $\alpha_1,\alpha_2$ and $\alpha_3$. We shall label
these by an index $\nu$. The corresponding Clebsch-Gordan coefficients
are defined as
\begin{equation*}
F^{(\nu)\, j_3}_f(\alpha_3, x_3) =: \sum_{j_2,j_1} \int_\mathbb{R}
\ud x_2 \ud x_1 \clebsh{\alpha_3}{x_3}^{(\nu)\, j_3}_{j_2j_1}
f^{j_2j_1} (x_2,x_1)\ .
\end{equation*}
In order to construct these coefficients, we introduce the following
products
\begin{equation}
D^{(\nu);\epsilon}_{\tau\sigma}(x_i;\alpha_i) = (-1)^{\nu(\tau+1)(\sigma+1)}
\frac{S_{\tau+\nu+1}(\z_{21}+\epsilon)}{{S_\tau}(\z_{21}+\epsilon +\a_{21})}
\frac{S_{\nu+\sigma+1}(\z_{23}+\epsilon)}{S_{\sigma}(\z_{23}+\epsilon+\a_{23})}
\frac{S_{\tau+\sigma+\nu}(\z_{13}-\epsilon)}{S_{\tau+\sigma+1}(\z_{13}-\epsilon+\a_{13})},
\end{equation}
where $\z_{ij}$ and $\a_{ij}$ in the same way as in the $\Uqtwo$ case.
In addition, we have introduced a parameter $\epsilon$ that will serve
as a regulator in products of Clebsch-Gordan coefficients later on, just
as in the case of $\Uqtwo$. The Clebsch-Gordan maps are now obtained as
\begin{equation} \label{eq:clebschosp}
 \clebsh{\alpha_3}{x_3}_{j_2j_1}^{(\nu)\, j_3}
= \sum_{\tau,\sigma} (-1)^{(j_1+\sigma)(j_2+\tau+\sigma)}
(-|\rho|)^{\nu(1-(j_1-j_2)^2) - j_1j_2}
\delta_{j_1+j_2+\nu,j_3}
{\mathcal N}^{1/2} D^{(\nu)}_{\tau\sigma}(x_i;\alpha_i)\ .
\end{equation}
The normalizing factor ${\mathcal N}^{1/2}$ is the square root of
the ${\mathcal N}$ we defined in eq.\ \eqref{Norm}.
Regularization is understood whenever it is necessary. If we
remove the regulator $\epsilon$, we obtain the Clebsch-Gordan
coefficients
\begin{equation*}
 \clebsh{\alpha_3}{x_3}^{(\nu)\, j_3} _{j_2j_1} =
 \lim_{\epsilon\rightarrow0} \left( {\clebsh{\alpha_3}{x_3}_\epsilon}
 \right)^{(\nu)\, j_3}_{j_2j_1}.
\end{equation*}
The intertwiner properties and orthogonality relations for these
Clebsch-Gordan coefficients are established following the same steps
as in the case of $\Uqtwo$. Our discussion in the subsequent section
will therefore focus on equations containing additional signs and on
the final results.

\subsection{Intertwiner property}

The Clebsh-Gordan coefficients satisfy the intertwining properties for
$X = K,v^{(\pm)}$
\begin{align*}
 \pi_{\alpha_3}(X)^i_{\phantom{i}{j_3}} \delta^{j_2}_j \delta^{j_1}_k \clebsh{\alpha_3}{x_3}^{j_3}_{{j_2}{j_1}} = \delta^i_{j_3} (\pi_{\alpha_2}\otimes\pi_{\alpha_1})\Delta^t(X)^{{j_2}\phantom{j}{j_1}}_{\phantom{{j_2}}j\phantom{{j_1}}k} \clebsh{\alpha_3}{x_3}^{j_3}_{{j_2}{j_1}}.
\end{align*}
The transpose on the right hand side is defined with respect to the
scalar product \eqref{eq:sp}. All these equations may be checked by
direct computations. For $X=K$ the analysis is identical to the one
outlined in section 3.1. So, let us proceed to $X=v^{(+)}$ right
away. When written out in components, our basic intertwining relation
reads
\begin{align*}
 (v_{\alpha_3}^{(+)})^1_{\phantom{1}0} \clebsh{\alpha_3}{x_3}^{(0)\, 0}_{00} &= \Delta_{10}^t(v^{(+)})^{0\phantom{0}1}_{\phantom{0}0\phantom{1}0} \clebsh{\alpha_3}{x_3}^{(0)\,1}_{01} + \Delta_{10}^t(v^{(+)})^{1\phantom{0}0}_{\phantom{1}0\phantom{0}0} \clebsh{\alpha_3}{x_3}^{(0)\,1}_{10},
 \\[1mm]
 (v_{\alpha_3}^{(+)})^0_{\phantom{0}1} \clebsh{\alpha_3}{x_3}^{(0)\,1}_{01} &=  \Delta_{10}^t(v^{(+)})^{0\phantom{0}0}_{\phantom{0}0\phantom{0}1} \clebsh{\alpha_3}{x_3}^{(0)\,0}_{00} + \Delta_{10}^t(v^{(+)})^{1\phantom{0}1}_{\phantom{1}0\phantom{1}1} \clebsh{\alpha_3}{x_3}^{(0)\,0}_{11},
 \\[1mm]
 (v_{\alpha_3}^{(+)})^0_{\phantom{0}1} \clebsh{\alpha_3}{x_3}^{(0)\,1}_{10} &= -\Delta_{10}^t(v^{(+)})^{1\phantom{1}1}_{\phantom{1}1\phantom{1}0} \clebsh{\alpha_3}{x_3}^{(0)\,0}_{11} + \Delta_{10}^t(v^{(+)})^{0\phantom{1}0}_{\phantom{0}1\phantom{0}0} \clebsh{\alpha_3}{x_3}^{(0)\,0}_{00},
 \\[1mm]
 (v_{\alpha_3}^{(+)})^1_{\phantom{1}0} \clebsh{\alpha_3}{x_3}^{(0)\,0}_{11} &= \Delta_{10}^t(v^{(+)})^{0\phantom{1}1}_{\phantom{0}1\phantom{1}1} \clebsh{\alpha_3}{x_3}^{(0)\,1}_{01} - \Delta_{10}^t(v^{(+)})^{1\phantom{1}0}_{\phantom{1}1\phantom{0}1} \clebsh{\alpha_3}{x_3}^{(0)\,1}_{10}.
\end{align*}
For the second set of Clebsch-Gordan coefficients we find,
\begin{align*}
 (v_{\alpha_3}^{(+)})^0_{\phantom{0}1} \clebsh{\alpha_3}{x_3}^{(1)\, 1}_{11} &= \Delta_{10}^t(v^{(+)})^{0\phantom{1}1}_{\phantom{0}1\phantom{1}1} \clebsh{\alpha_3}{x_3}^{(1)\, 0}_{01} - \Delta_{10}^t(v^{(+)})^{1\phantom{1}0}_{\phantom{1}1\phantom{0}1} \clebsh{\alpha_3}{x_3}^{(1)\, 0}_{10},
 \\[2mm]
 (v_{\alpha_3}^{(+)})^0_{\phantom{0}1} \clebsh{\alpha_3}{x_3}^{(1)\, 1}_{00} &= \Delta_{10}^t(v^{(+)})^{0\phantom{0}1}_{\phantom{0}0\phantom{1}0} \clebsh{\alpha_3}{x_3}^{(1)\, 0}_{01} + \Delta_{10}^t(v^{(+)})^{1\phantom{0}0}_{\phantom{1}0\phantom{0}0} \clebsh{\alpha_3}{x_3}^{(1)\, 0}_{10},
 \\[2mm]
 (v_{\alpha_3}^{(+)})^1_{\phantom{1}0} \clebsh{\alpha_3}{x_3}^{(1)\, 0}_{01} &= \Delta_{10}^t(v^{(+)})^{0\phantom{0}0}_{\phantom{0}0\phantom{0}1} \clebsh{\alpha_3}{x_3}^{(1)\, 1}_{00} + \Delta_{10}^t(v^{(+)})^{1\phantom{0}1}_{\phantom{1}0\phantom{1}1} \clebsh{\alpha_3}{x_3}^{(1)\, 1}_{11},
 \\[2mm]
 (v_{\alpha_3}^{(+)})^1_{\phantom{1}0} \clebsh{\alpha_3}{x_3}^{(1)\, 0}_{10} &= -\Delta_{10}^t(v^{(+)})^{1\phantom{1}1}_{\phantom{1}1\phantom{1}0} \clebsh{\alpha_3}{x_3}^{(1)\, 1}_{11} + \Delta_{10}^t(v^{(+)})^{0\phantom{1}0}_{\phantom{0}1\phantom{0}0} \clebsh{\alpha_3}{x_3}^{(1)\, 1}_{00},
\end{align*}
As in the nonsupersymetric case one may employ the identities
\begin{align*}
 T^{i\ub}_x \, \frac{S_1(-ix+a_1)}{S_1(-ix+a_2)} &=
 \frac{[-ix + a_1]_1}{[-ix + a_2]_1} \frac{S_0(-ix+a_1)}{S_0(-ix+a_2)}\, T^{i\ub}_x ,\\[2mm]
 T^{i\ub}_x \, \frac{S_0(-ix+a_1)}{S_0(-ix+a_2)} &= \frac{[-ix + a_1]_0}{[-ix + a_2]_0}
 \frac{S_1(-ix+a_1)}{S_1(-ix+a_2)} \, T^{i\ub}_x ,\\[2mm]
 T^{i\ub}_x \, \frac{S_0(-ix+a_1)}{S_1(-ix+a_2)} &= -i \frac{q^{\frac{1}{2}} -
 q^{-\frac{1}{2}}}{q^{\frac{1}{2}} + q^{-\frac{1}{2}}} \frac{[-ix + a_1]_0}{[-ix + a_2]_1}
 \frac{S_1(-ix+a_1)}{S_0(-ix+a_2)}\, T^{i\ub}_x .
\end{align*}
to check that all the intertwining relation for $X = v^{(+)}$ is satisfied. The
same steps are carried out to discuss $X = v^{(-)}$. Details are left to the
reader.

\subsection{Orthogonality and Completeness}

The most difficult part in the analysis of the Clebsch-Gordan coefficients is
once again concerning their orthonormality relations. The intertwining relations
we have established in the previous subsection guarantee that
\begin{eqnarray}\nonumber
& & \sum_{j_2,j_2}\int_\mathbb{R}\ud x_2 \ud x_1
\overline{\left[^{\alpha_3}_{x_3} \phantom{}^{\alpha_2}_{x_2}
\phantom{}^{\alpha_1}_{x_1}\right]}^{(\nu),j_3}_{j_2j_1}
\left[^{\beta_3}_{y_3} \phantom{}^{\alpha_2}_{x_2} \phantom{}^{\alpha_1}_{x_1}
\right]_{k_2 k_1}^{(\mu),i_3} \lambda^{j_2k_2} \lambda^{j_1 k_1}
= \\[2mm]
&  & \hspace*{0.5cm} =
32 \sqrt{\rho}^{(-1)^{(\nu+1)}} \sum_\sigma (-1)^{(\nu+1)(\sigma+1)} |S_\sigma(2\alpha_3)|^{-2}
\delta_{\nu,\mu} \lambda^{j_3,i_3} \delta(\bbeta_3 - \balpha_3)
\delta(y_3 - x_3), \label{orthoosp}
\end{eqnarray}
up to an overall factor. This normalization is established with the help
of a set of integral identities which follow from a supersymmetric version
of the star-triangle identity, see Appendix B.2. In particular one uses
$$\sum_{\tau,\sigma} \int dx_2 dx_1 (-1)^{(\rho+\nu+\mu)\tau}\left(\tilde D^{(\mu)\, \epsilon}_{\tau\sigma}\right)^\ast \,
D^{(\nu) \, \epsilon}_{\tau(\sigma+\rho)} = \frac{16 (-1)^{\rho\nu}}{
|S_{\rho+1}(2\balpha_3)|^{2}} \delta_{\mu,\nu}\delta(\bbeta_3-\balpha_3)\delta(y_3-x_3),
$$
where we introduced
$$ D_{\tau\sigma}^{(\nu)\,\epsilon} = D_{\tau\sigma}^{(\nu) \, \epsilon}
(x_3,x_2,x_1;\alpha_3,\alpha_2,\alpha_1)
\quad , \quad
\tilde D_{\tau\sigma}^{(\nu)\,\epsilon} =
D_{\tau\sigma}^{(\nu) \, \epsilon}
(y_3,x_2,x_1;\beta_3,\alpha_2,\alpha_1)\ . $$
Since the analogous computation for $\Uqtwo$ was described in great detail
in section 3.2 we can leave the derivation of eq.\ \eqref{orthoosp} as an
exercise.

\section{The Racah-Wigner coefficients for $\Uqosp$}

The definition and computation of the Racah-Wigner coefficients for $\Uqosp$
proceeds very much along the same lines as for $\Uqtwo$, see section 4.
After giving a precise definition in the Racah-Wigner coefficients, we will
state an explicit formula. It resembles the one for $\Uqtwo$, see eq.\
\eqref{RWsltwo}, except that all special functions carry an additional
label $\nu \in \{0,1\}$.

As in the case of $\Uqtwo$ be begin by defining the following two
maps for the decomposition of triple tensor products,
\begin{eqnarray} \label{Phitosp}
 \left( \Phi^t_{\alpha_t} \left[{\alpha_3 \atop \alpha_4}{\alpha_2\atop\alpha_1}\right]^{\nu_1\nu_2}_\epsilon\right)^i_{\phantom{i}jkl}(x_4; x_i) &
 = & \sum_n \int \ud x_t \left[ {\alpha_4 \atop x_4}{\alpha_t \atop x_t}{\alpha_1 \atop x_1}\right]^{(\nu_1)\, i}_{\epsilon \phantom{i} n l} \left[ {\alpha_t \atop x_t}{\alpha_3 \atop x_3}{\alpha_2 \atop x_2}\right]^{(\nu_2)\, n}_{\epsilon \phantom{\alpha}jk}\\[2mm]
 \left( \Phi^s_{\alpha_s} \left[{\alpha_3 \atop \alpha_4}{\alpha_2\atop\alpha_1}\right]^{\nu_3\nu_4}_\epsilon \right)^i_{\phantom{i}jkl}(x_4; x_i) & = & \sum_m \int \ud x_s  \left[ {\alpha_4 \atop x_4}{\alpha_3 \atop x_3}{\alpha_s \atop x_s}\right]^{(\nu_3)\, i}_{\epsilon \phantom{i}jm} \left[ {\alpha_s \atop x_s}{\alpha_2 \atop x_2}{\alpha_1 \atop x_1}\right]^{(\nu_4)\, m}_{\epsilon \phantom{m}kl}\ .
\label{Phisosp}
\end{eqnarray}
From these two maps we can compute the Racah-Wigner coefficients through the
usual prescription
\begin{eqnarray}\label{6jintosp}
 \left(\left\{ {\alpha_1 \atop \alpha_2} {\alpha_3\atop\alpha_4} | {\alpha_s \atop \alpha_t}\right\}_\ub\right)^{\nu_3\nu_4}_{\nu_1\nu_2}
  & = & \\[2mm]
  & & \hspace*{-4.3cm} = \lim_{\epsilon\to0} \sum_{jklm}
 \int \ud\alpha_4\int \ud^4 x
 \left( \left( \Phi^t_{\alpha_t} \left[{\alpha_3 \atop \alpha'_4}{\alpha_2\atop\alpha_1}\right]^{\nu_1\nu_2}_\epsilon\right)^{m}_{jkl}
 (x'_4; x_i) \right)^* \left( \Phi^s_{\alpha_s} \left[{\alpha_3 \atop
 \alpha_4}{\alpha_2\atop\alpha_1}\right]^{\nu_3\nu_4}_\epsilon\right)^n_{jkl}(x_4; x_i)
\nonumber
\end{eqnarray}
where the integration measure is $d^4 x = \prod_{i=1}^4 dx_i$. After integration and
summation, the right hand side turns out to be independent of $\a_4',x'_4$ and $n$.
Using the explicit formulas for the regularized Clebsch-Gordan maps along with
our knowledge of poles and residues of the special functions $S_\nu$, see Appendix
B.2, we can perform the integrations with the help of Cauchy's integral formula to
obtain,
\begin{eqnarray} \label{RWosp}
   \left(\left\{ {\alpha_1 \atop \alpha_2} {\alpha_3\atop\alpha_4} | {\alpha_s \atop \alpha_t}\right\}_\ub\right)^{\nu_3\nu_4}_{\nu_1\nu_2} & &  \ = \ \delta_{\sum_i \nu_i=0 \ \mbox{\it mod\/}\  2}\  (-1)^{\nu_2\nu_3+\nu_4} \
  \frac{S_{\nu_4}(a_4) S_{\nu_1}(a_1)}{S_{\nu_2}(a_2)S_{\nu_3}(a_3) } \times \\[2mm]
 & & \ \hspace*{-4.3cm} \times \ \int_{i\mathbb{R}} \ud t \sum_{\nu=0}^1
  (-1)^{\nu(\nu_2+\nu_4)}
  \frac{S_{1+\nu+\nu_4} (u_4+t)S_{1+\nu+\nu_4}(\tilde u_4+t)S_{1+\nu+\nu_3}( u_3+t)S_{1+\nu+\nu_3}(\tilde u_3+t)}
  {S_{\nu+\nu_2+\nu_3}(u_{23} +t)S_{\nu+\nu_2+\nu_3}(\tilde u_{23}+t)
  S_{\nu}(2\alpha_s+t)S_{\nu}(Q+t)}\ . \nonumber
\end{eqnarray}
All the variables $a_i$ and $u_i, \tilde u_i$ etc.\ where defined in
section 4. Note that they are associated to the four vertices which in turn
correspond to the indices $\nu_i$. In this form, our result appears as a
natural extension of the expression \eqref{RWsltwo} for the Racah-Wigner
coefficients of $\Uqtwo$. The sum over $\nu$ accompanies the integral
over t. The shift $\nu \rightarrow \nu+1$ in the index of $S_\nu$ appears
for those $S_\nu$ that we decided to write into the numerator. The
parameters $\nu_i$ are placed such that they mimic the arguments of
the $S_\nu$. Unfortunately, we do not have a simple rule to
explain the sign factor, but of course it comes out of the
calculation as stated.

\section{Comparison with fusing matrix of Liouville theory}

As we outlined in the introduction, the Racah-Wigner coefficients for
the self-dual series of representations we considered here coincide with
the Fusing matrices of (supersymmetric) Liouville theory, at least up to
some normalization dependent prefactors. For the case of $\Uqtwo$ this
was shown by Teschner in \cite{Teschner:2001rv,Teschner:2003en}. Entries
of the fusing matrix of $N=1$ supersymmetric Liouville theory were
computed in \cite{Hadasz:2007wi} and we are now going to compare these with the
Racah-Wigner symbols \eqref{RWosp} of $\Uqosp$, after a short review
of the non-supersymmetric theory.

\subsection{Liouville field theory and $\Uqtwo$}

The fusion matrix of Liouville field theory can be obtained by calculating
the exchange relations for the chiral operators in the scalar field
representation \cite{Teschner:2001rv,Teschner:2003en} (see also
\cite{Gervais:1993fh} for an earlier construction). To spell out this
result we need to introduce some notation.

The Verma module ${\cal V}_{\Delta}$ of the Virasoro algebra of the
highest weight $\Delta$ and the central charge $c$ is defined as a
free vector space generated by all vectors of the form
\begin{equation}
\nu_{\Delta,MK}
\; = \;
L_{-m_j}\ldots L_{-m_1}\nu_{\Delta}\,,
\end{equation}
where
\(
m_j \geqslant \ldots \geqslant m_2 \geqslant m_1,\hspace{10pt} m_r\in \mathbb{N}
\)
and $\nu_{\Delta}$ is the highest weight state with respect to the Virasoro algebra,
\begin{equation}
L_0\nu_{\Delta}=\Delta\nu_{\Delta},
\hskip 5mm
L_m\nu_{\Delta}  =0,
\hskip 5mm
m > 0.
\end{equation}
The chiral vertex operator,
$$
V\chiral{\Delta_2}{\Delta_3}{\Delta_1}(z)\;:\;{\cal V}_{\Delta_1} \;\to\;{\cal V}_{\Delta_3},
$$
is a linear map parameterized by the conformal weight of the ``intermediate''
module ${\cal V}_{\Delta_2}$ and defined by the commutation relations
\begin{equation}
\label{commutation:chiral:V}
\big[L_n, V\chiral{\Delta_2}{\Delta_3}{\Delta_1}(z)\big] \; = \; z^n\big(z\partial_z +(n+1)\Delta_2\big)V\chiral{\Delta_2}{\Delta_3}{\Delta_1}(z),
\end{equation}
the form of the  Virasoro algebra and a normalization
\begin{equation}
V\chiral{\Delta_2}{\Delta_3}{\Delta_1}(z)\nu_{\Delta_1} \; =
\; z^{\Delta_3-\Delta_2-\Delta_1}\left(\hbox{\boldmath $1$} +
{\cal O}(z)\right)\nu_{\Delta_3}, \hskip 1cm z \to 0.
\end{equation}
With the notion of the chiral vertex operator at hand we can define a four-point conformal block,
\begin{equation}
{\cal F}_{\Delta_s}\!\left[^{\Delta_3\: \Delta_2}_{\Delta_4\: \Delta_1}\right]\!(z)
\; = \;
\left\langle \nu_{\Delta_4},
V\chiral{\Delta_3}{\Delta_4}{\Delta_s}(1)V\chiral{\Delta_2}{\Delta_s}{\Delta_1}(z)\nu_{\Delta_1}
\right\rangle,
\end{equation}
where $\langle\,\cdot\,,\,\cdot\,\rangle$ is the usual hermitian, bilinear form on
${\cal V}_{\Delta},$ which is uniquely characterized by the conditions $L_n^\dag = L_{-n},\ $
and $\langle\nu_\Delta,\nu_\Delta\rangle = 1.$ Finally, the Liouville fusion matrix (or
monodromy of conformal block) is defined as an integral kernel appearing in the relation
\begin{equation}
{\cal F}_{\Delta_s}\!\left[^{\Delta_3\: \Delta_2}_{\Delta_4\: \Delta_1}\right]\!(z)
\; = \;
\int_{{\frac{\mathbb{Q}}{2}} + i\mathbb{R^+}} \ud\alpha_t\
F_{\alpha_s\,\alpha_t}\!\left[^{\alpha_3\: \alpha_2}_{\alpha_4\: \alpha_1}\right]
{\cal F}_{\Delta_t}\!\left[^{\Delta_1\: \Delta_2}_{\Delta_4\: \Delta_3}\right]\!(1-z),
\end{equation}
with
\(
\Delta_i = \alpha_i(Q-\alpha_i).
\)
It appears to be difficult to derive the form of the fusion matrix directly
from its definition. However, there exists a simple relation (which we shall
formulate explicitly below) between the fusion matrix and the braiding matrix
of the Virasoro chiral vertex operators, i.e.\ the integral kernel appearing in
the formula
\begin{equation}
\label{brading:matrix:V}
V\chiral{\Delta_3}{\Delta_4}{\Delta_s}(z_2)V\chiral{\Delta_2}{\Delta_s}{\Delta_1}(z_1)\
=
\int_{{\frac{\mathbb{Q}}{2}} + i\mathbb{R^+}} \ud\alpha_u\
B_{\alpha_s\,\alpha_u}\!\left[^{\alpha_3\: \alpha_2}_{\alpha_4\: \alpha_1}\right]
V\chiral{\Delta_2}{\Delta_4}{\Delta_u}(z_1)V\chiral{\Delta_3}{\Delta_u}{\Delta_1}(z_2).
\end{equation}
The latter can be derived by direct calculations of the exchange relation of
chiral vertex operators in the free field representation
\cite{Teschner:2001rv,Teschner:2003en}.

Let the hermitian operators ${\sf p,q},$ with $[{\sf p,q}] = -\frac{i}{2},$
act on the Hilbert space $L^2({\mathbb R}).$ Denote by ${\cal F}$ the Fock
space generated by action of negative modes of the algebra
\[
[{\sf a}_m,{\sf a}_n] = \frac12m\delta_{m,-n},\;\; m,n \in {\mathbb Z}\setminus\{0\},
\hskip 1cm
{\sf a}_m^\dag = {\sf a}_{-m},
\]
on the vacuum $\Omega,$ where ${\sf a}_m\Omega = 0,\; m > 0.$ On the Hilbert space
${\cal H} = L^2({\mathbb R})\otimes {\cal F}$ there exists a well known representation
of the Virasoro algebra with the central charge $c = 1+ 6Q^2,$ given by,
\begin{eqnarray}
\nonumber
L_m({\sf p}) & = & \sum\limits_{n\neq 0,m}{\sf a}_{n}{\sf a}_{m-n} +
(2{\sf p} + imQ){\sf a}_m, \hskip 1cm m \neq 0,
\\[-8pt]
\label{Virasoro:free}
\\[-8pt]
\nonumber
L_0({\sf p}) & = & 2\sum\limits_{n=1}^\infty {\sf a}_{n}{\sf a}_{-n} + {\sf p}^2 + \frac14 Q^2.
\end{eqnarray}
Each state of the form $|\nu_p\rangle \equiv |p\rangle\otimes\Omega,$ with
${\sf p}|p\rangle = p|p\rangle,$ is of highest weight with respect to
the algebra (\ref{Virasoro:free}) and satisfies
\[
L_0({\sf p}) |\nu_p\rangle = \Delta(p)|\nu_p\rangle,
\hskip 1cm
\Delta(p) = p^2 + \frac14 Q^2 = \alpha(Q-\alpha),\;\;
\alpha = \frac{Q}{2} + ip.
\]
Acting on such a state with $L_{-n}({\sf p})$ one generates a
Virasoro Verma module ${\cal V}_{\Delta(p)}.$

The normal ordered exponentials
\begin{equation}
\label{ordered:exponentials}
{\sf E}^\alpha(x) =
{\rm e}^{\alpha{\sf q}}\,
{\rm e}^{2\alpha\varphi_<(x)}\,
{\rm e}^{2\alpha x {\sf p}}\,
{\rm e}^{2\alpha\varphi_>(x)}\,
{\rm e}^{\alpha{\sf q}}
\end{equation}
where
\[
\varphi_<(x) = -i \sum\limits_{n =1}^\infty\frac{{\sf a}_{-n}}{n}\,{\rm e}^{inx},
\hskip 1cm
\varphi_<(x) = i \sum\limits_{n = 0}^\infty\frac{{\sf a}_n}{n}\,{\rm e}^{-inx},
\]
together with the screening charge
\[
{\sf Q}(x) = \int_{x}^{x+2\pi}\,{\sf E}^{\rm b}(y)dy
\]
serve as building block for a chiral field
\begin{equation}
\label{chiral:g}
{\sf g}^\alpha_s(x) = {\sf E}^\alpha(x)\big({\sf Q}(x)\big)^s.
\end{equation}
Commutation relations of the field (\ref{chiral:g})  with the Virasoro algebra
generators are of the form
\begin{equation}
\label{commutation}
[L_n({\sf p}),{\sf g}^\alpha_s(x)] = {\rm e}^{inx}\left(-i\frac{d}{dx} +n\alpha(Q-\alpha)\right){\sf g}^\alpha_s(x)
\end{equation}
and coincide with (\ref{commutation:chiral:V}) (with $\Delta(p)$ substituted for
$\Delta_2$) for $w$ chiral vertex operator transformed from the complex $z$ plane
to the zero-time slice of an infinite cylinder.  It is also easy to check that
${\sf g}^\alpha_s(x)$ maps vectors form ${\cal V}_{p_1}$ to ${\cal V}_{q}$ with
$q = p_1 -i(\alpha + bs).$ The field  (\ref{chiral:g}) therefore yields a model
for a family of (unnormalized) vertex operators with $\Delta_2 = \Delta(p)$
and arbitrary (thanks to a possibility of adjusting the value of $s$) $\Delta_1
= \Delta(p_1)$ and $\Delta_3 = \Delta(q).$

Suppose there exists an exchange relation
\begin{equation}
\label{exchange:gg}
{\sf g}^{\alpha_2}_{s_2}(x_2){\sf g}^{\alpha_1}_{s_1}(x_1)
=
\int dt_1dt_2\
B(\alpha_i|s_i,t_i)\,
{\sf g}^{\alpha_1}_{t_1}(x_1){\sf g}^{\alpha_2}_{t_2}(x_2).
\end{equation}
Using
\[
{\sf E}^{\alpha}(x){\sf E}^{\beta}(y) = {\rm e}^{-2\pi i\, {\rm sign}(x-y)}\,
{\sf E}^{\beta}(y){\sf E}^{\alpha}(x),
\]
and a clever representation of the screening charges in term of a Weyl type
operators it is possible to normal order both sides of (\ref{exchange:gg}).
Schematically
\begin{eqnarray*}
{\sf g}^{\alpha_2}_{s_2}(x_2){\sf g}^{\alpha_1}_{s_1}(x_1)
& = &
{\sf E}^{\alpha_2}(x_2){\sf E}^{\alpha_1}(x_1)A({\sf x})P_{2,1}({\sf p,t}),
\\
{\sf g}^{\alpha_1}_{t_1}(x_1){\sf g}^{\alpha_2}_{t_2}(x_2)
& = &
{\sf E}^{\alpha_2}(x_2){\sf E}^{\alpha_1}(x_1)A({\sf x})P_{1,2}({\sf p,t}),
\end{eqnarray*}
where the functions $A, P_{1,2}$ and $P_{2,1}$ are explicitly known and the
operators ${\sf x, p, t}$ satisfy commutation relations
\[
[{\sf p,x}] = -\frac{i}{2},
\hskip 1cm
[{\sf p,t}] =[{\sf t,x}] = 0.
\]
Upon acting on a common eigenstate of ${\sf p}$ and ${\sf t}$ the formula
(\ref{exchange:gg}) thus boils down to a relation involving just functions
of eigenvalues of these operators and the integral kernel $B$ we are after.
The special function $G_{\rm b}$ appears in this formula thanks to identities
of the form
\[
{\rm e}^{{\rm b}{\sf x}}\left(1+{\rm e}^{-2\pi{\rm b}{\sf p}}\right){\rm e}^{-{\rm b}{\sf x}}
=
G_{\rm b}\left({\textstyle\frac{Q}{2}}+i{\sf p}\right){\rm e}^{2{\rm b}{\sf x}}G^{-1}_{\rm b}\left({\textstyle\frac{Q}{2}}+i{\sf p}\right)
=
{\rm e}^{2{\rm b}{\sf x}}G_{\rm b}\left({\textstyle\frac{Q}{2}}+i{\sf p}+ {\rm b}\right)G^{-1}_{\rm b}\left({\textstyle\frac{Q}{2}}+i{\sf p}\right)
\]
which also allow to calculate an arbitrary power of the screening charge ${\sf Q}.$

To calculate the braiding matrix of the normalized vertex operators, appearing
in eq.\ (\ref{brading:matrix:V}), one needs to determine the matrix element of
the chiral field ${\sf g}^{\alpha}_{s}(1)$ between highest weight states of the
Verma modules ${\cal V}_{\Delta(p_1)}$ and ${\cal V}_{\Delta(q)}.$ This was
achieved in \cite{Teschner:2001rv,Teschner:2003en} by deriving and solving a
pair of difference equations for this matrix element, which follow from
considering the four-point correlation function involving degenerate field
${\sf E}^{-\frac{\rm b}{2}}(x).$

Finally, a relation between braiding matrix and the fusion matrix can be derived
by considering a sequence of ``moves'' including braiding of generic vertices,
``elementary'' braiding of a generic vertex with the vertex acting on the vacuum
Verma module (with $\Delta = 0$) and use of a state-operator correspondence
\cite{Teschner:2003en} (see also \cite{Hadasz:2004cm} for a more detailed
explanation). It reads
\begin{equation}
F_{\alpha_s\,\alpha_t}\!\left[^{\alpha_3\: \alpha_2}_{\alpha_4\: \alpha_1}\right]
=
B_{\alpha_s\,\alpha_t}\!\left[^{\alpha_3\: \alpha_1}_{\alpha_4\: \alpha_2}\right].
\end{equation}
The resulting form of the Liouville fusion matrix, to be found for instance
in \cite{Ponsot:2003ju} section 3.5, coincides with eq.\ (\ref{RWsltwo}) up
to a factor due to a different normalizations of chiral vertex operators and
Clebsh-Gordan coefficients.

\subsection{Neveu-Schwarz sector of the ${\cal N}=1$ superconformal theory}
The Neveu-Schwarz (or NS for short) supermodule ${\cal V}_\Delta$
of the highest weight $\Delta$ and the central charge $c$
is a  free vector space spanned by vectors of the form
\begin{equation}
\label{basis}
\nu_{\Delta,MK}
\; = \;
L_{-M}G_{-K} \nu_{\Delta}
\; \equiv \;
L_{-m_j}\ldots L_{-m_1}G_{-k_i}\ldots G_{-k_1}\nu_{\Delta}\,,
\end{equation}
where
 $K = \{k_1,k_2,\ldots,k_i\}$ and
 $M = \{m_1,m_2,\ldots,m_j\}$ are
arbitrary ordered sets of  indices
$$
k_i > \ldots > k_2 > k_1 ,
\hspace{10pt} k_s\in \mathbb{N}-{1\over 2}, \hspace{20pt}
m_j \geqslant \ldots \geqslant m_2 \geqslant m_1,\hspace{10pt} m_r\in \mathbb{N}.
$$
Here $\nu_{\Delta}$ is the highest weight state of to the NS
algebra,
\begin{eqnarray}
\label{NS}
\nonumber
\left[L_m,L_n\right] & = & (m-n)L_{m+n} +\frac{c}{12}m\left(m^2-1\right)\delta_{m+n},
\\
\left[L_m,G_k\right] & = &\frac{m-2k}{2}G_{m+k},
\\
\nonumber
\left\{G_k,G_l\right\} & = & 2L_{k+l} + \frac{c}{3}\left(k^2 -\frac14\right)\delta_{k+l},
\hskip 1cm
c = \frac32 + 3Q^2,
\end{eqnarray}
even with respect to the
the fermion parity operator $(-1)^F$ defined by relations
\begin{eqnarray}
\nonumber
[(-1)^F, L_m]&=&\{(-1)^F,G_k\}=0.
\end{eqnarray}
The NS module is thus a direct sum of an even and an odd (with respect to $(-1)^F$) subspaces
\[
{\cal V}_\Delta = {\cal V}_\Delta^{\rm e}\oplus {\cal V}_\Delta^{\rm o}.
\]
This ${\mathbb Z}_2$ grading reflects itself in a parity structure of the chiral vertex
operators: we define them as two families of even,
\[
V^{\rm e}\chiral{\Delta_2}{\Delta_3}{\Delta_1}(z): \; {\cal V}_{\Delta_1}^{\eta}\; \to \; {\cal V}_{\Delta_3}^{\eta},
\hskip 5mm
V^{\rm e}\chiral{\hskip -2pt *\Delta_2}{\Delta_3}{\Delta_1}(z): \; {\cal V}_{\Delta_1}^{\eta}\; \to \; {\cal V}_{\Delta_3}^{\eta},
\hskip 1cm
\eta = {\rm e, o}
\]
and two families of odd linear maps
\[
V^{\rm o}\chiral{\Delta_2}{\Delta_3}{\Delta_1}(z): \; {\cal V}_{\Delta_1}^{\eta}\; \to \; {\cal V}_{\Delta_3}^{\bar\eta},
\hskip 5mm
V^{\rm o}\chiral{\hskip -2pt *\Delta_2}{\Delta_3}{\Delta_1}(z): \; {\cal V}_{\Delta_1}^{\eta}\; \to \; {\cal V}_{\Delta_3}^{\bar\eta},
\hskip 1cm
\bar{\rm e} = {\rm o},\ \bar{\rm o} = {\rm e},
\]
uniquely specified by the (anti)commutation relations
(here $ \underline{\hspace*{6pt}}\Delta_2$ stands either
for  $\Delta_2$ or for $*\Delta_2$)
\begin{eqnarray}
\nonumber
\big[L_m,V^{\eta}\chiral{\Delta_2}{\Delta_3}{\Delta_1}(z)\big]
& = &
z^m\left(z \partial_z + (m+1) \Delta_2 \right)V^{\eta}\chiral{\Delta_2}{\Delta_3}{\Delta_1}(z),
\\[4pt]
\nonumber
\big[L_m,V^{\eta}\chiral{\hskip -2pt *\Delta_2}{\Delta_3}{\Delta_1}(z)\big]
& = &
z^m\left(z \partial_z + (m+1)\left(\Delta_2 + \textstyle \frac{1}{2}\right)\right)V^{\eta}\chiral{\hskip -2pt *\Delta_2}{\Delta_3}{\Delta_1}(z),
\\[-6pt]
\\[-7pt]
\nonumber
\big[G_k,V^{\rm e}\chiral{\hskip -2pt \underline{\hspace*{3pt}}\Delta_2}{\Delta_3}{\Delta_1}(z)\big]
& = &
z^{k+ \frac{1}{2}}\, V^{\rm o}\chiral{\hskip -2pt \underline{\hspace*{3pt}}\Delta_2}{\Delta_3}{\Delta_1}(z),
\\[4pt]
\nonumber
\big\{G_k,V^{\rm o}\chiral{\hskip -2pt \underline{\hspace*{3pt}}\Delta_2}{\Delta_3}{\Delta_1}(z)\big\}
& = &
z^{k - \frac{1}{2}} \left( z \partial_z + \Delta_2 (2k+1) \right)
V^{\rm e}\chiral{\hskip -2pt \underline{\hspace*{3pt}}\Delta_2}{\Delta_3}{\Delta_1}(z),
\end{eqnarray}
and appropriate normalization conditions. We are thus in a position to define four even
\begin{eqnarray}
\label{even:blocks}
{\cal F}_{\Delta_s}^{\rm e}\!\left[^{\underline{\hspace*{4pt}}\Delta_3\ \underline{\hspace*{4pt}}\Delta_2}_{\hspace*{4pt}\Delta_4\ \hspace*{4pt}\Delta_1}\right]\!(z)
& = &
\left\langle\nu_{\Delta_4},
V^{\rm e}\chiral{\hskip -2pt \underline{\hspace*{3pt}}\Delta_3}{\Delta_4}{\Delta_s}(1)
V^{\rm e}\chiral{\hskip -2pt \underline{\hspace*{3pt}}\Delta_2}{\Delta_s}{\Delta_1}(z)
\nu_{\Delta_1}
\right\rangle
\end{eqnarray}
and four odd conformal blocks,
\begin{eqnarray}
\label{odd:blocks}
{\cal F}_{\Delta_s}^{\rm o}\!\left[^{\underline{\hspace*{4pt}}\Delta_3\ \underline{\hspace*{4pt}}\Delta_2}_{\hspace*{4pt}\Delta_4\ \hspace*{4pt}\Delta_1}\right]\!(z)
& = &
\left\langle\nu_{\Delta_4},
V^{\rm o}\chiral{\hskip -2pt \underline{\hspace*{3pt}}\Delta_3}{\Delta_4}{\Delta_s}(1)
V^{\rm o}\chiral{\hskip -2pt \underline{\hspace*{3pt}}\Delta_2}{\Delta_s}{\Delta_1}(z)
\nu_{\Delta_1}
\right\rangle.
\end{eqnarray}
As in the Liouville case we define the fusion matrices $F$ by the relation
\begin{eqnarray}
\label{NS:fusion}
{\cal F}_{\Delta_s}^{\rm \eta}\!\left[^{\underline{\hspace*{4pt}}\Delta_3\ \underline{\hspace*{4pt}}\Delta_2}_{\hspace*{4pt}\Delta_4\ \hspace*{4pt}\Delta_1}\right]\!(z)
& = &
\int\limits_{{\frac{\mathbb{Q}}{2}} + i\mathbb{R^+}}\hskip -10pt \ud\alpha_t\
\sum\limits_{\rho = {\rm e,o}}
F_{\alpha_s\alpha_t}\!\left[^{\underline{\hspace*{4pt}}\alpha_3\: \underline{\hspace*{4pt}}\alpha_2}_{\hspace*{4pt}\alpha_4\: \hspace*{4pt}\alpha_1}\right]^{\eta}{}_{\rho}\
{\cal F}_{\Delta_s}^{\rm \rho}\!\left[^{\underline{\hspace*{4pt}}\Delta_1\ \underline{\hspace*{4pt}}\Delta_2}_{\hspace*{4pt}\Delta_4\ \hspace*{4pt}\Delta_3}\right]\!(1-z).
\end{eqnarray}
Calculation of the braiding matrices above, given in  \cite{Chorazkiewicz:2008es},
is parallel to the calculation in the Liouville case and we shall not present any
of its details here, referring the interested reader to the original paper. To
relate the findings of that paper to the form of the 6j symbols given in
(\ref{RWosp}) let us start by observing that formulae (5.10), (4.61) and (4.53)
from \cite{Chorazkiewicz:2008es} give
\begin{eqnarray*}
&&
\hskip -1.5cm
{F}_{\alpha_s\alpha_t}\left[^{\alpha_3\ \alpha_2}_{\alpha_4\ \alpha_1}\right]^{\rm e}{}_{\rm e}
\\[4pt]
& = &
\frac{
\Gamma_1(\alpha_t+\alpha_4-\alpha_1)\Gamma_1(\bar\alpha_t+\alpha_4-\alpha_1)
\Gamma_1(\alpha_t+\bar\alpha_4-\alpha_1)\Gamma_1(\bar\alpha_t+\bar\alpha_4-\alpha_1)
}{
\Gamma_1(\alpha_s+\alpha_4-\alpha_3)\Gamma_1(\bar\alpha_s+\alpha_4-\alpha_3)
\Gamma_1(\alpha_s+\bar\alpha_4-\alpha_3)\Gamma_1(\bar\alpha_s+\bar\alpha_4-\alpha_3)}
\\[4pt]
& \times &
\frac{
\Gamma_1(\alpha_t+\alpha_2-\alpha_3)\Gamma_1(\bar\alpha_t+\alpha_2-\alpha_3)
\Gamma_1(\alpha_t+\bar\alpha_2-\alpha_3)\Gamma_1(\bar\alpha_t+\bar\alpha_2-\alpha_3)
}{
\Gamma_1(\alpha_s+\alpha_2-\alpha_1)\Gamma_1(\bar\alpha_s+\alpha_2-\alpha_1)
\Gamma_1(\alpha_s+\bar\alpha_2-\alpha_1)\Gamma_1(\bar\alpha_s+\bar\alpha_2-\alpha_1)}
\\[4pt]
& \times &
\frac14
\frac{
\Gamma_1(2\alpha_s)\Gamma_1(2\bar\alpha_s)
}{
\Gamma_1(Q-2\alpha_t)\Gamma_1(2\alpha_t-Q)}\
\int\limits_{i{\mathbb R}}
\frac{dt}{i}\
{J}_{\alpha_s\alpha_t}\left[^{\alpha_3\ \alpha_1}_{\alpha_4\ \alpha_2}\right]
\end{eqnarray*}
where
\begin{eqnarray*}
{J}_{\alpha_s\alpha_t}\left[^{\alpha_3\ \alpha_1}_{\alpha_4\ \alpha_2}\right]
& = &
\sum\limits_{\nu=0}^1
\frac{
S_\nu(\alpha_2+t)S_\nu(\bar\alpha_2+t)
S_\nu(\bar\alpha_4 -\alpha_3+\alpha_1+t)S_\nu(\alpha_4 -\alpha_3+\alpha_1+t)
}{
S_\nu(\alpha_t+\bar\alpha_3+t)S_\nu(\bar\alpha_t+\bar\alpha_3+t)
S_\nu(\alpha_s+\alpha_1+t)S_\nu(\bar\alpha_s+\alpha_1+t)}.
\end{eqnarray*}
It was proven in \cite{Hadasz:2007wi} that this function enjoys the following
symmetry property
\begin{eqnarray*}
{F}_{\alpha_s\alpha_t}\left[^{\alpha_3\ \alpha_2}_{\alpha_4\ \alpha_1}\right]^{\rm e}{}_{\rm e}
& = &
{F}_{\alpha_s\alpha_t}\left[_{\alpha_3\ \alpha_2}^{\bar\alpha_4\ \alpha_1}\right]^{\rm e}{}_{\rm e}
\end{eqnarray*}
so that we have
\begin{eqnarray*}
&&
\hskip -1.5cm
{F}_{\alpha_s\alpha_t}\left[^{\alpha_3\ \alpha_2}_{\alpha_4\ \alpha_1}\right]^{\rm e}{}_{\rm e}
\\[4pt]
& = &
\frac{
\Gamma_1(\alpha_t+\alpha_3-\alpha_2)\Gamma_1(\bar\alpha_t+\alpha_3-\alpha_2)
\Gamma_1(\alpha_t+\bar\alpha_3-\alpha_2)\Gamma_1(\bar\alpha_t+\bar\alpha_3-\alpha_2)
}{
\Gamma_1(\alpha_s+\alpha_3-\bar\alpha_4)\Gamma_1(\bar\alpha_s+\alpha_3-\bar\alpha_4)
\Gamma_1(\alpha_s+\bar\alpha_3-\bar\alpha_4)\Gamma_1(\bar\alpha_s+\bar\alpha_3-\bar\alpha_4)}
\\[4pt]
& \times &
\frac{
\Gamma_1(\alpha_t+\alpha_1-\bar\alpha_4)\Gamma_1(\bar\alpha_t+\alpha_1-\bar\alpha_4)
\Gamma_1(\alpha_t+\bar\alpha_1-\bar\alpha_4)\Gamma_1(\bar\alpha_t+\bar\alpha_1-\bar\alpha_4)
}{
\Gamma_1(\alpha_s+\alpha_1-\alpha_2)\Gamma_1(\bar\alpha_s+\alpha_1-\alpha_2)
\Gamma_1(\alpha_s+\bar\alpha_1-\alpha_2)\Gamma_1(\bar\alpha_s+\bar\alpha_1-\alpha_2)}
\\[4pt]
& \times &
\frac14
\frac{
\Gamma_1(2\alpha_s)\Gamma_1(2\bar\alpha_s)
}{
\Gamma_1(Q-2\alpha_t)\Gamma_1(2\alpha_t-Q)}\
\int\limits_{i{\mathbb R}}
\frac{dt}{i}\
{J}_{\alpha_s\alpha_t}\left[_{\alpha_3\ \alpha_1}^{\bar\alpha_4\ \alpha_2}\right]
\end{eqnarray*}
where, after a shift of the integration variable $t \to t+\alpha_s-\alpha_2,$
\begin{equation}
\label{susy:current:final}
\int\limits_{i{\mathbb R}} \frac{dt}{i}\
{J}_{\alpha_s\alpha_t}\left[_{\alpha_3\ \alpha_1}^{\bar\alpha_4\ \alpha_2}\right]
=
\int\limits_{i{\mathbb R}}\frac{dt}{i}\ \sum\limits_{\nu=0}^1
\frac{S_\nu(u_4+t)S_\nu(\tilde u_4 +t) S_\nu(u_3+t)S_\nu(\tilde u_3 +t)
}{
S_\nu(u_{23}+t)S_\nu(\tilde u_{23} +t) S_\nu(2\alpha_s+t)S_\nu(Q+t)}
\end{equation}
with the variables $u_i, \tilde u_i$ etc.\ defined in section 4.
Comparing eq.\ (\ref{susy:current:final}) with eq.\ (\ref{RWosp}) we
thus see that
\[
F\left[^{\alpha_3\ \alpha_2}_{\alpha_4\ \alpha_1}\right]^{\rm e}{}_{\rm e}
\; \propto\;
\left(\left\{ {\alpha_1 \atop \alpha_2} {\alpha_3\atop\alpha_4} | {\alpha_s \atop \alpha_t}\right\}_\ub\right)^{1\,1}_{1\,1},
\]
with the proportionality constant again due to a different normalization
of chiral vertex operators and 6j symbols.

Repeating the same calculation for the other components of the fusion
matrix given by eq.\ (5.10) from \cite{Chorazkiewicz:2008es} we obtain
\[
F\left[^{\alpha_3\ \alpha_2}_{\alpha_4\ \alpha_1}\right]^{\rm e}{}_{\rm o}
\; \propto\;
\left(\left\{ {\alpha_1 \atop \alpha_2} {\alpha_3\atop\alpha_4} | {\alpha_s \atop \alpha_t}\right\}_\ub\right)^{1\,1}_{0\,0}
\]
and
\[
F\left[^{\alpha_3\ \alpha_2}_{\alpha_4\ \alpha_1}\right]^{\rm o}{}_{\rm e}
\; \propto\;
\left(\left\{ {\alpha_1 \atop \alpha_2} {\alpha_3\atop\alpha_4} | {\alpha_s \atop \alpha_t}\right\}_\ub\right)^{0\,0}_{1\,1},
\hskip 1cm
F\left[^{\alpha_3\ \alpha_2}_{\alpha_4\ \alpha_1}\right]^{\rm o}{}_{\rm o}
\; \propto\;
\left(\left\{ {\alpha_1 \atop \alpha_2} {\alpha_3\atop\alpha_4} | {\alpha_s \atop \alpha_t}\right\}_\ub\right)^{0\,0}_{0\,0}.
\]
Similarly, components of the fusion matrix given by formula (5.11) from
\cite{Chorazkiewicz:2008es}, as well as the remaining eight components of
the fusion matric appearing in (\ref{NS:fusion}) not explicitly given in
that paper, are expressible in terms of the 6j symbols (\ref{RWosp}).

\section{Conclusions}

In this work we have constructed and studied a set of infinite dimensional
self-dual representations of the quantum deformed algebra $\Uqosp$. In
particular, we computed the Clebsch-Gordan coefficients \eqref{eq:clebschosp}
for the decomposition of tensor products of self-dual representations and the
associated Racah-Wigner coefficients \eqref{RWosp}. All these data were
built out of a pair of special functions $\Phi^\pm_\ub(z)$. In our analysis
we employed a number of beautiful integral identities for these functions,
see section 6 and appendix B.2. These mimic corresponding identities
satisfied by the quantum dilogarithm, only that all integrations are now
accompanied by a summation of discrete indices $\nu=0,1$.

There are a number of issues that would be interesting to address. To begin
with, we have only computed the fusing matrix for the Neveu-Schwarz sector
of $N=1$ Liouville field theory. It would certainly be important to include
matrix elements when some of the fields are taken from the Ramond-sector.
Representations from the Ramond sector of the field theory should be
associated with another series of self-dual representations of $\Uqosp$.
Once these representations have been identified one needs to extend the
above analysis to products within such an enlarged class of representations. 
In field theory, at least some elements of the extended fusing matrix are 
known. Since their form is not very different from what we encountered in 
the Neveu-Schwarz sector, the entire analysis is likely to rest on the 
set of integral identities we stated and used above.

Another interesting direction is to extend the number of supersymmetries.
In stepping from $N=1$ to $N=2$ supersymmetric Liouville theory, we must
replace $\Uqosp$ by $\mathcal{U}_q(osp(2|2))= \mathcal{U}_q(sl(1|2))$.
Even though this is probably the most relevant case, it would be possible
to continue and study the entire series $\mathcal{U}_q(osp(N|2))$. 
Finally, let us also mention that there exists an intriguing duality
between the 6j symbols for finite dimensional representations of 
$\Uqtwo$ and $\Uqosp$, see \cite{Saleur:1989gj,Ennes:1997kx}. It would 
be very interesting to extend this duality to the self-dual series. We 
shall return to these issues in forthcoming publications.    
\smallskip

\noindent
{\bf Acknowledgements:} We wish to thank Paulina Suchanek, J\"org Teschner,
Grigory Vartanov and Edward Witten for discussions and useful comments. This
work was supported in part by the GRK 1670 ``Mathematics inspired by Quantum
Field and String Theory''. The work of LH was supported by the Polish Science
Centre (NCN) grant 2011/01/B/ST1/01302 ``Conformal blocks in two-dimensional
field theories''.

\begin{appendix}
\section{Some special functions}

This appendix contains definitions and a short list of
important properties for the special functions that appear
in the main text. The first subsections deals with those
functions that arise in the context of $\Uqtwo$ while the
second subsection is tailored towards out discussion of
$\Uqosp$. Derivations of some of the identities can be found at
many places in the literature.

\subsection{Special functions for $\Uqtwo$}\label{def_specfunc}

The basic building block for all objects that appear in the context
of the quantum algebra $\Uqtwo$ is Barnes' double Gamma function. For
$\mathfrak{Re}x>0$ it admits an integral representation
\begin{eqnarray*}
 \log \Gamma_\ub(x) = \int_0^\infty \frac{\ud t}{t}
 \left[ \frac{e^{-x t} - e^{-\frac{Q}{2} t}}{(1 - e^{-t \ub})
 (1 - e^{-\frac{t}{\ub}})} - \frac{\left( \frac{Q}{2}-x\right)^2}{2 e^t}
 - \frac{\frac{Q}{2} - x}{t} \right] ,
\end{eqnarray*}
where $Q = \ub + \frac{1}{\ub}$. One can analytically continue $\Gamma_\ub$
to a meromorphic function defined on the entire complex plane
$\mathbb{C}$. The most important property of $\Gamma_\ub$ is its
behavior with respect to shifts by $\ub^\pm$,
\begin{equation}
 \Gamma_\ub(x+\ub) = \frac{\sqrt{2\pi} \ub^{\ub x-\frac{1}{2}}}{\Gamma_\ub(bx)}
 \Gamma_\ub(x)\quad , \quad
 \Gamma_\ub(x+\ub^{-1}) = \frac{\sqrt{2\pi} \ub^{-\frac{\ub}{x}+\frac{1}{2}}}
 {\Gamma_\ub(\frac{x}{\ub})} \Gamma_\ub(x)\ .
\end{equation}
These shift equation allows us to calculate residues of the poles of
$\Gamma_\ub$. When $x\to0$, for instance, one finds
\begin{equation}\label{residue}
 \Gamma_\ub(x)= \frac{\Gamma_\ub(Q)}{2\pi x} + O(1).
\end{equation}
From Barnes' double Gamma function we can build two other important
special functions,
\begin{align}
 S_\ub(x) &= \frac{\Gamma_\ub(x)}{\Gamma_\ub(Q-x)}, \\[2mm]
 G_\ub(x) &= e^{-\frac{i\pi}{2} x(Q-x)} S_\ub(x).\label{definition_G}
\end{align}
We shall often refer to the function $S_\ub$ as double sine function.
It is related to Faddeev's quantum dilogarithm through,
$$  \Phi_\ub(x) =\ A G_\ub^{-1}(-ix + \frac{Q}{2}),
$$
where
\begin{equation} \label{eq:A}
 A \, = \, e^{-i\pi(1-4 c_\ub^2)\slash 12}\quad , \quad  c_\ub = i Q/2\ .
\end{equation}
The $S_\ub$ function is meromorphic with poles and zeros in
\begin{align*}
  S_\ub(x) = 0 &\Leftrightarrow x = Q + n b + m b^{-1}, \quad n,m \in \mathbb{Z}_{\geq0}\ , \\[2mm]
  S_\ub(x)^{-1} = 0 &\Leftrightarrow x = -n b -m b^{-1}, \quad n,m \in \mathbb{Z}_{\geq0}\ .
\end{align*}
From its definition and the shift property of Barnes' double Gamma
function it is easy to derive the following shift and reflection
properties of $G_\ub$,
\begin{align}
 &G_\ub(x+\ub) = (1-e^{2 \pi i \ub x}) G_\ub(x)\ ,\label{shift} \\[2mm]
 &G_\ub(x) G_\ub(Q-x) = e^{\pi i x(x-Q)}\ .\label{reflection}
\end{align}
We also need to the asymptotic behavior of the function $G_\ub$ along
the imaginary axis,
\begin{equation}\begin{split}\label{asymptotic}
 G_\ub(x) \ \sim &\  1\ , \qquad \qquad \mathfrak{Im} x \to +\infty,\\[2mm]
 G_\ub(x) \sim & \ e^{i\pi x(x-Q)}\, , \quad \mathfrak{Im}x \to -\infty.
\end{split}
\end{equation}

\subsection{Special functions for $\Uqosp$}

In discussing the representation theory of the quantum superalgebra
$\Uqosp$ we need the following additional special functions
\begin{align*}
 \Gamma_1(x) & = \Gamma_\NS(x) = \Gamma_\ub\left(\frac{x}{2}\right) \Gamma_\ub
 \left(\frac{x+Q}{2}\right),\\[2mm]
 \Gamma_0(x) & = \Gamma_\R(x) = \Gamma_\ub\left(\frac{x+\ub}{2}\right) \Gamma_\ub
 \left(\frac{x+\ub^{-1}}{2}\right).
\end{align*}
Furthermore, let us define
\begin{equation}
\begin{array}{rlrl}
 S_1(x)&\!= S_\NS(x) = \frac{\Gamma_\NS(x)}{\Gamma_\NS(Q-x)} , \quad &
G_1(x)&\!= G_\NS(x) = \zeta_0 e^{-\frac{i\pi}{4} x(Q-x)} S_\NS(x), \\[2mm]
  S_0(x)&\!= S_\R(x) = \frac{\Gamma_\R(x)}{\Gamma_\R(Q-x)}, \quad &
 G_0(x)&\!= G_\R(x) = e^{-\frac{i\pi}{4}} \zeta_0 e^{-\frac{i\pi}{4}
 x(Q-x)} S_\R(x)
\end{array}
\end{equation}
where $\zeta_0 = \exp(-i\pi Q^2/8)$. The functions $S_\nu$ are related to
the supersymmetric analogue of Faddeev's quantum dilogarithm through
$$
\Phi^\nu_\ub(x) =\ A^2 G_\nu^{-1}(-ix + \frac{Q}{2}),
$$
with a constant $A$ as defined in eq.\ \eqref{eq:A}.
As for $S_\ub$, the functions $S_0(x)$ and $S_1(x)$ are meromorphic with poles and zeros in
\begin{align*}\
  S_0(x) = 0 &\Leftrightarrow x = Q + n b + m b^{-1}, \quad n,m \in \mathbb{Z}_{\geq0}, m+n\in2\mathbb{Z}+1, \\
  S_1(x) = 0 &\Leftrightarrow x = Q + n b + m b^{-1}, \quad n,m \in \mathbb{Z}_{\geq0}, m+n\in2\mathbb{Z},\\
  S_0(x)^{-1} = 0 &\Leftrightarrow x = -n b -m b^{-1}, \quad n,m \in \mathbb{Z}_{\geq0}, m+n\in2\mathbb{Z}+1, \\
  S_1(x)^{-1} = 0 &\Leftrightarrow x = -n b -m b^{-1}, \quad n,m \in \mathbb{Z}_{\geq0}, m+n\in2\mathbb{Z}.
\end{align*}
As in the previous subsection, we want to state the shift and reflection properties of the functions $G_1$
and $G_0$,
\begin{align}
  G_\nu(x+\b^{\pm1}) &= (1-(-1)^\nu e^{ \pi i \ub^{\pm1} x}) G_{\nu+1}(x),
  \label{shift_super}\\[2mm]
  \label{reflection_super}
  G_\nu(x) G_\nu(Q-x) &= e^{\frac{i\pi}{2}(\nu-1)} \zeta_0^2
  e^{\frac{\pi i}{2} x(x-Q)}\ .
\end{align}
Asymptotically, the functions $G_1$ and $G_0$ behave as
\begin{align}
\label{asymptotic_superNS}
 G_\nu(x) &\sim 1\ , \qquad\qquad\qquad\quad\quad
 \ \ \mathfrak{Im}x\to +\infty\ ,\\[2mm]
 G_\nu(x) &\sim e^{\frac{i\pi}{2}(\nu-1)} \zeta_0^2 e^{\frac{i\pi}{2}
 x(x-Q)}\ , \,\quad \mathfrak{Im}x\to -\infty\ .
\end{align}

\section{Integral identities}

For our proof of the orthogonality relations of Clebsch Gordan coefficients
we need a number of integral formulas for the special functions discussed in
the previous section. We shall state these identities here. In both subsections
we shall start with the star triangle relations and then deduce a number of
simpler integral identities.

\subsection{Integral identities for $\Uqtwo$}

The most complex identity we need in the main text is the
following star triangle relation for double sine function,
\begin{equation*}
 \int \frac{\ud x}{i} \prod_{i=1}^3 S_\ub(x+a_i)S_\ub(-x+b_i)
 = \prod_{i,j=1}^3 S_\ub(a_i+b_j)\ ,
\end{equation*}
which holds provided that the arguments satisfy the balancing
condition
\begin{equation*}
 \sum^3_{i=1}(a_i+b_i) = Q.
\end{equation*}
A proof can be found e.g.\ in \cite{kashaev}. Here, we will
only state the necessary results. The star triangle relation
can be reduced to the Saalsch\"utz summation formula
\begin{align*}
& \frac{1}{i} \int_{-i\infty}^{i\infty} \ud\tau e^{2\pi i \tau Q} \frac{G_\ub(\tau + a) G_\ub(\tau + b) G_\ub(\tau + c)}{G_\ub(\tau + a + b + c - d + Q) G_\ub(\tau + Q) G_\ub(\tau + d)} = \\[2mm]
&= e^{i\pi d(Q-d)} G_\ub(a) G_\ub(b) G_\ub(c) \frac{G_\ub(Q + b - d) G_\ub(Q+c-d) G_\ub(Q+a-d)}{G_\ub(Q + b +c -d) G_\ub(Q+a+c-d) G_\ub(Q+a+b-d)}.
\end{align*}
A useful consequence of the Saalsch\"utz summation formula can
be obtained by taking the limit $c\to i\infty$
\begin{align*}\label{saalschutz}
& \int_{-i\infty}^{i\infty} \frac{\ud\tau}{i} e^{2\pi i \tau Q}
\frac{G_\ub(\tau + a) G_\ub(\tau + b)}{G_\ub(\tau + d) G_\ub(\tau + Q)} =\\[2mm]
&=e^{i\pi d(Q-d)} G_\ub(a) G_\ub(b) G_\ub(Q + b - d)
\frac{G_\ub(Q+a-d)}{G_\ub(Q+a+b-d)}.
\end{align*}
Also, by taking the additional limits $a\to-i\infty$, $d\to-i\infty$ with
$a-d+Q$ fixed one may derive the well known Ramanujan summation formula
\begin{equation} \label{Ramanujan}
 \int_{-i\infty}^{i\infty} \frac{\ud\tau}{i} e^{2 \pi i \tau \beta} \frac{G_\ub(\tau + \alpha)}{G_\ub(\tau + Q)} = \frac{G_\ub(\alpha) G_\ub(\beta)}{G_\ub(\alpha+\beta)},
\end{equation}
which holds for arbitrary $\alpha = a-d+Q$ and $\beta=b$. Ramanujan's summation
formula is a five-term (pentagon) identity. In may be considered a quantization
of the familiar Rogers five-term identity satisfied by dilogarithms. In fact,
the function $G_\ub$ that was used throughout most of this test is closely
related to Faddeev's quantum dilogarithm $\Phi_\ub$ which we introduced in the
introduction, see eq.\ \eqref{dilog}.

\subsection{Integral identities for $\Uqosp$}

In the supersymmetric case, the star triangle relations take the following
form
\begin{align*}
\sum_{\nu=0,1} (-1)^{\nu(1+\sum_i(\nu_i+\mu_i))/2}
\int \frac{\ud x}{i} \prod_{i=1}^3 S_{\nu+\nu_i}(x+a_i)
S_{1+\nu+\mu_i}(-x+b_i) &= 2 \prod_{i,j=1}^3 S_{\nu_i+\mu_i}(a_i+b_j),
\end{align*}
with
\begin{align}
\sum_i (\nu_i +\mu_i) = 1 \ \mbox{\it mod\/} \ 2
\end{align}
and the balancing condition
\begin{equation*}
 \sum^3_{i=1}(a_i+b_i) = Q\ .
\end{equation*}
From these equations one can get 16 ``supersymmetric'' analogues of
the Saalsch\"utz summation formula, some of which are stated with
proofs for instance in \cite{Hadasz:2007wi}. As in the non-supersymmetric
case, taking the limit $d\to i\infty$ leads to the reduced formulae
\begin{equation*}\begin{split}
 &\sum_{\sigma=0,1} \int_{-i\infty}^{i\infty} \frac{\ud \tau}{i}
  e^{i\pi\tau Q}
  \frac{G_{\sigma+\rho_a}(\tau+a)G_{\sigma+\rho_\b}(\tau+b)}
  {G_{\sigma+\rho_c}(\tau+c)G_{1+\sigma}(\tau+Q)} = \\
&\qquad = 2 i^{1-\rho_c} \zeta_0^{-3} e^{\frac{i\pi}{2} c (Q-c)}
\frac{G_{\rho_a}(a)G_{\rho_b}(b)G_{1+\rho_a-\rho_c}(Q+a-c)
G_{1+\rho_b-\rho_c}(Q+b-c)}{G_{\rho_a+\rho_b-\rho_c}(Q+a+b-c)}.
\end{split}\end{equation*}
where $\zeta_0 = \exp(-i\pi Q^2/8)$ is the same constant factor
as before. From these identities one can easily obtain a system
of four equations that generalize Ramanujan's formula
\eqref{Ramanujan} to the supersymmetric case,
\begin{equation}
\sum_{\sigma=0,1} \int_{-i\infty}^{i\infty} \frac{\ud\tau}{i} (-1)^{\rho_\beta\sigma}
e^{ \pi i \tau \beta} \frac{G_{\sigma+\rho_\alpha}(\tau + \alpha)}{G_{\sigma+1}
(\tau + Q)} = 2 \zeta_0^{-1} \frac{G_{\rho_\alpha}(\alpha)
G_{1+\rho_\beta}(\beta)}{G_{\rho_\alpha+\rho_\beta}(\alpha+\beta)}
\end{equation}
The notations are the same as in section B.1. The last identity is is supersymmetric
version of the pentagon identity for Faddeev's quantum dilogarithm.

\section{Removing the regulator}

Lets consider the distribution
\begin{equation*}
 D(x,\xi_-) = \lim_{\epsilon\rightarrow0} \frac{S_\ub(2\epsilon+x) S_\ub(-\xi_- -x) S_\ub(2\epsilon + \xi_-) }{S_\ub(4\epsilon)}.
\end{equation*}
One wants to show that the following holds
\begin{equation*}
 D(x,\xi_-) = \delta(x)\delta(i\xi_-).
\end{equation*}
In order to do that, it is sufficient to independently integrate over the first or the second variable and establish that in both cases the result is proportional to the appropriate delta function.

For any test function $f(y,x)$, one can consider:
\begin{align*}
 \int \frac{\ud x}{i}\frac{\ud y}{i} f(y,x) D(x,y) &= \lim_{\epsilon\rightarrow0} \int \frac{\ud x}{i}\frac{\ud y}{i} f(y,x)  \frac{S_\ub(2\epsilon+x) S_\ub(-\xi_- -x) S_\ub(2\epsilon + \xi_-) }{S_\ub(4\epsilon)} \\
 &= \lim_{\epsilon\rightarrow0} (2\pi)^{-2} \int \frac{\ud x}{i}\frac{\ud y}{i} f(y,x) \frac{\epsilon}{(\epsilon+x)(x+y)(\epsilon+y)} = \\
 &= \lim_{\epsilon\rightarrow0} (2\pi)^{-2} \int \frac{\ud x}{i}\frac{\ud y}{i} f(\epsilon y,\epsilon x) \frac{1}{(1+x)(x+y)(1+y)} = \\
 &= \left( (2\pi)^{-2} \int \frac{\ud x}{i}\frac{\ud y}{i} \frac{1}{(1+x)(x+y)(1+y)}\right) f(0,0),
\end{align*}
and one only needs to fix the multiplicative constant.

\subsubsection*{Calculation for $\xi_-$}

One wants to evaluate the expression
\begin{equation*}
 \int \frac{\ud \xi_-}{i} f(\xi_-,x) D(x,\xi_-) = f(0,0)\delta(ix) .
\end{equation*}
As a test function, lets choose $f(\xi,x) = \exp(-i\pi \xi (x+2a))$, for $a \in i\mathbb{R}\backslash\{0\}$ and $x \in i\mathbb{R}$. To begin with, let us define
\begin{align*}
 \mu &= -x + 2\epsilon,\\
 \tau &= -(\xi_-+2\epsilon),
\end{align*}
so that one has
\begin{align*}
 &\int \frac{\ud\xi_-}{i} e^{-i\pi\xi_- (x+2a)} S_\ub(2\epsilon+\xi_-)S_\ub(-\xi_--x) =\\
 &= \int\frac{\ud\tau}{i} e^{i\pi(\tau+2\epsilon)(x+2a)} \frac{S_\ub(\tau+\mu)}{S_\ub(\tau+Q)} =\\
 &= e^{2i\pi\epsilon (x+2a)} \int\frac{\ud\tau}{i} e^{i\pi\tau (x+2a)} e^{-i\pi\tau\mu}e^{\frac{i\pi}{2}(Q-\mu)\mu} \frac{G_\ub(\tau+\mu)}{G_\ub(\tau+Q)} =\\
 &= e^{2i\pi\epsilon (x+2a)} e^{\frac{i\pi}{2}(Q-\mu)\mu} \int\frac{\ud\tau}{i} e^{2i\pi\tau(\frac{x+2a-\mu}{2})} \frac{G_\ub(\tau+\mu)}{G_\ub(\tau+Q)} =\\
 &= e^{2i\pi\epsilon (x+2a)} e^{\frac{i\pi}{2}(Q-\mu)\mu} \frac{G_\ub(\mu) G_\ub(\frac{x+2a-\mu}{2})}{G_\ub(\frac{x+2a+\mu}{2})} = \\
 &= e^{2i\pi\epsilon (x+2a)} e^{\frac{i\pi}{2}(Q-\mu)\mu} \frac{G_\ub(2\epsilon-x) G_\ub(x+a-\epsilon)}{G_\ub(a+\epsilon)},
\end{align*}
where one used Ramanujan summation formula. Therefore
\begin{align*}
 &\int \frac{\ud \xi_-}{i} f(\xi_-,x) D(x,\xi_-) = \int\frac{\ud\xi_-}{i}\lim_{\epsilon\rightarrow0} e^{-i\pi \xi (x+2a)} \frac{S_\ub(2\epsilon+x) S_\ub(-\xi_- -x) S_\ub(2\epsilon + \xi_-) }{S_\ub(4\epsilon)} = \\
&=\lim_{\epsilon\rightarrow0} \frac{S_\ub(2\epsilon+x)}{S_\ub(4\epsilon)} \int  \frac{\ud\xi_-}{i} e^{-i\pi\xi_- (x+2a)} S_\ub(2\epsilon+\xi_-)S_\ub(-\xi_--x) =\\
&= e^{-\frac{i\pi}{2}(Q+x)x} \lim_{\epsilon\rightarrow0} \frac{S_\ub(2\epsilon+x)}{S_\ub(4\epsilon)}\frac{G_\ub(2\epsilon-x) G_\ub(x+a-\epsilon)}{G_\ub(a+\epsilon)} =\\
&= e^{-i\pi x^2} \frac{ G_\ub(x+a)}{G_\ub(a)} \lim_{\epsilon\rightarrow0} \frac{G_\ub(2\epsilon+x)G_\ub(2\epsilon-x)}{G_\ub(4\epsilon)} = (*)
\end{align*}
Since $\lim_{x\rightarrow0}xG_\ub(x)=\frac{1}{2\pi}$ and it is known that
\begin{equation*}
 \lim_{\epsilon\rightarrow0} \frac{2\epsilon}{\pi(4\epsilon^2-x^2)} = \delta(ix),
\end{equation*}
one obtains eventually
\begin{align*}
 (*) &= e^{-i\pi x^2} \frac{ G_\ub(x+a)}{G_\ub(a)} \delta(ix) = \delta(ix).
\end{align*}

\subsubsection*{Calculation for $x$}

Now one can repeat the above procedure for integration over $x$, i.e. show that
\begin{equation*}
 \int \frac{\ud x}{i} g(\xi_-,x) D(x,\xi_-) = g(0,0) \delta(i\xi_-).
\end{equation*}
Lets take a different test function $g$ s.t.
\begin{equation*}
 g(x,\xi_-) = e^{-i\pi x(\xi_-+2a)},
\end{equation*}
where $A\in i\mathbb{R}\backslash\{0\}$. Then define
\begin{align*}
 \mu &= -\xi_- + 2\epsilon,\\
 \tau &= -(x+2\epsilon),
\end{align*}
so that one has
\begin{align*}
 &\int \frac{\ud x}{i} e^{-i\pi x(\xi_-+2a)} S_\ub(2\epsilon+x)S_\ub(2\epsilon-\xi_--x) =\\
 &= e^{2i\pi\epsilon(\xi_-+2a)} e^{\frac{i\pi}{2}(Q-\mu)\mu} \int\frac{\ud\tau}{i} e^{2i\pi\tau(\frac{\xi_-+2a-\mu}{2})} \frac{G_\ub(\tau+\mu)}{G_\ub(\tau+Q)} =\\
 &= e^{2i\pi\epsilon(\xi_-+2a)} e^{\frac{i\pi}{2}(Q-\mu)\mu} \frac{G_\ub(2\epsilon-\xi_-) G_\ub(\xi_-+a-\epsilon)}{G_\ub(a+\epsilon)},
\end{align*}
where one again used Ramanujan summation formula. Therefore
\begin{align*}
 &\int\frac{\ud x}{i} g(\xi_-,x) D(x,\xi_-) = \lim_{\epsilon\rightarrow0} \frac{S_\ub(2\epsilon+\xi_-)}{S_\ub(4\epsilon)} \int  \frac{\ud x}{i} e^{-i\pi x(\xi_-+2a)} S_\ub(2\epsilon+x)S_\ub(-\xi_--x) =\\
&= e^{-i\pi \xi_-^2} \lim_{\epsilon\rightarrow0} \frac{G_\ub(2\epsilon+\xi_-)G_\ub(2\epsilon-\xi_-)}{G_\ub(4\epsilon)} = e^{-i\pi \xi_-^2} \delta(i\xi_-) = \delta(\xi_-).
\end{align*}

Putting the two results together one concludes that
\begin{equation*}
 D(x,\xi_-) = \delta(ix)\delta(i\xi_-).
\end{equation*}

\end{appendix}

\end{document}